\documentclass[twocolumn,showpacs,showkeys,preprintnumbers,amsmath]{revtex4}

\usepackage{amsfonts}

\DeclareFontFamily{U}{msa}{}
\DeclareFontShape{U}{msa}{m}{n}
    { <-> msam10}{}
\DeclareSymbolFont{AMSa}{U}{msa}{m}{n}

\DeclareFontFamily{U}{msb}{}
\DeclareFontShape{U}{msb}{m}{n}
     { <-> msbm10}{}
\DeclareSymbolFont{AMSb}{U}{msb}{m}{n}

\DeclareFontFamily{U}{euf}{}
\DeclareFontShape{U}{euf}{m}{n}
    { <-> eufm10}{}
\DeclareFontShape{U}{euf}{b}{n}
    { <-> eufb10}{}

\DeclareFontFamily{U}{eus}{}
\DeclareFontShape{U}{eus}{m}{n}
    { <-> eusm10}{}
\DeclareFontShape{U}{eus}{b}{n}
    { <-> eusb10}{}

\DeclareFontFamily{U}{eur}{}
\DeclareFontShape{U}{eur}{m}{n}
    { <-> eurm10}{}
\DeclareFontShape{U}{eur}{b}{n}
    { <-> eurb10}{}

\DeclareMathAlphabet{\matheurm}{U}{eur}{m}{n}
\DeclareMathAlphabet{\matheuf}{U}{euf}{m}{n}
\DeclareMathAlphabet{\matheurmbf}{U}{eur}{b}{n}
\DeclareMathAlphabet{\matheuscr}{U}{eus}{m}{n}
\DeclareMathAlphabet{\mathsfsl}{OT1}{cmss}{m}{sl}
\DeclareMathAlphabet{\mathsf}{OT1}{cmss}{m}{n}
\DeclareFontShape{OT1}{cmr}{bx}{n}{ <-> cmbx10 }{}

\DeclareMathSymbol{\smallfrown}{\mathrel}{AMSa}{"61}
\DeclareMathSymbol{\subsetneq}{\mathrel}{AMSb}{"24}
\DeclareMathSymbol{\therefore}{\mathrel}{AMSa}{"29}

\newcommand{\concat}{{}^\smallfrown}

\newcommand{\famind}[1]{_{(#1)}}
\newcommand{\mathand}{\text{\ \&\ }}
\newcommand{\mathor}{\text{\normalfont\ or\ }}

\newcommand{\cnum}[1]{\matheurm #1}

\DeclareMathSymbol\subsetneqq{\mathbin}{AMSb}{"A4}
\DeclareMathSymbol\nsubseteq{\mathbin}{AMSb}{"AA}
\DeclareMathSymbol\compl{\mathord}{AMSb}{"73}
\DeclareMathSymbol\hbar{\mathord}{AMSb}{"7E}
\DeclareMathSymbol\gtrless{\mathbin}{AMSa}{"BF}
\DeclareMathSymbol\dotplus{\mathbin}{AMSa}{"75}
\DeclareMathSymbol\blacktriangleup{\mathord}{AMSa}{"4E}

{_{\substack{}{}}}

\newif\ifdisplaynum

\newcommand{\disptitle}[1]%
{\ifdisplaynum
  \hspace{-0.25em}{\sc #1}\hspace{0.5em}
 \else
  \noindent {\sc #1} \hspace{0.5em}
 \fi}

{\displaynumtrue
 \refstepcounter{equation}

 \begin{list}{}{%
  \setlength{\leftmargin}{0pt}}
  \item[](\theequation) \sl}%
{\end{list}}

\newenvironment{display*}%
{\displaynumfalse
 \begin{list}{}{%
  \setlength{\leftmargin}{0pt}}
  \item[]\sl}%
{\end{list}}

\newlength{\displayparindent}
{\begin{quote}
  \refstepcounter{equation}
  \dimen0=\leftmargin \advance \dimen0 by \parindent
  \makebox[0pt][r]{\makebox[\dimen0][l]{(\theequation)}} \sl}%
{\end{quote}}

\newenvironment{displist*}%
{\begin{quote}
  \sl}%
{\end{quote}}

\newcounter{bean}
{\begin{list}%
{(\arabic{bean})}%
{\usecounter{bean}\setlength{\rightmargin}{0pt}}}%
{\end{list}}

{\begin{list}%
{$\bullet$}%
{\setlength{\rightmargin}{0pt}}}%
{\end{list}}

\newlength{\gnat}
\newcommand{\midslash}[1]%
{\settowidth{\gnat}{$#1$}%
#1\hspace{-.5\gnat}\makebox[0pt]{/}%
\hspace{.5\gnat}}

\newcommand{\opneg}{{\boldsymbol{\neg}}}
\newcommand{\opand}{{\boldsymbol{\wedge}}}
\newcommand{\opor}{{\boldsymbol{\vee}}}
\newcommand{\opimplies}{{\boldsymbol{\rightarrow}}}
\newcommand{\opif}{{\boldsymbol{\leftarrow}}}
\newcommand{\opiff}{{\boldsymbol{\leftrightarrow}}}
\newcommand{\opin}{{\boldsymbol{\in}}}
\newcommand{\opmid}{{\boldsymbol{\mid}}}
\newcommand{\opeq}{{\boldsymbol{=}}}
\newcommand{\opforall}{{\boldsymbol{\forall}}}
\newcommand{\opexists}{{\boldsymbol{\exists}}}

\newcommand{\lcneg}{{\neg}}
\newcommand{\lcand}{{\wedge}}
\newcommand{\lcor}{{\vee}}
\newcommand{\lcimplies}{{\rightarrow}}
\newcommand{\lcif}{{\leftarrow}}
\newcommand{\lciff}{{\leftrightarrow}}

\newcommand{\llcneg}{{\,\neg\,}}
\newcommand{\llcand}{{\,\wedge\,}}
\newcommand{\llcor}{{\,\vee\,}}
\newcommand{\llcimplies}{{\,\rightarrow\,}}
\newcommand{\llcif}{{\,\leftarrow\,}}
\newcommand{\llciff}{{\,\leftrightarrow\,}}

\newcommand{\bv}[1]{{[\![#1]\!]}}
\newcommand{\bigbv}[1]{{\big[\!\!@,@,@,\big[#1\big]\!\!@,@,@,\big]}}

\newlength{\charwidth}
\newcommand{\slsh}[1]%
{#1\settowidth{\charwidth}{$#1$}\hspace{-.5\charwidth}\makebox[0pt]{$\slash$}%
\hspace{.5\charwidth}}

\newcommand{\tzf}{{\theory{ZF}}}
\newcommand{\tzfa}{{\theory{ZFA}}}
\newcommand{\tzfc}{{\theory{ZFC}}}
\newcommand{\tzfca}{{\theory{ZFCA}}}
\newcommand{\complet}[1]{{\overline{#1}}}

\newcommand{\visavis}{\emph{vis-\`a-vis\ }}

\newcommand{\Gcheck}{{\check\id}}

\newcommand{\Prob}{\operatorname{Prob}}
\newcommand{\tr}{\operatorname{tr}}
\renewcommand{\Re}{\operatorname{Re}}
\renewcommand{\Im}{\operatorname{Im}}

\newcommand{\vecsp}[1]{{\mathsf #1}}

\newcommand{\calcmd}[1]{{\mathcal #1}}

\newcommand{\spproj}{\proj{\mathstrut\ }}
\newcommand{\spbreve}{\breve{\mathstrut\ }}

\newcommand{\eqdef}{\overset{\mathrm{def}}{=}}
\newcommand{\iffdef}{\overset{\mathrm{def}}{\iff}}
\newcommand{\tiffdef}{$\iffdef$}

\newcommand{\image}{{}^{\scriptscriptstyle\rightarrow}}
\newcommand{\invimage}{{}^{\scriptscriptstyle\leftarrow}}

\newcommand{\qed}{\dimen0 = 1ex%
~\rule[-.55 \dimen0]{1.12 \dimen0}{2.25 \dimen0}%
\vspace{\baselineskip}}%

\newsavebox{\circb}
\savebox{\circb}(2.3,2){
\begin{picture}(1,1)(1,-1.25)
 \thinlines
 \put(0,0){\circle*{1.25}}
\end{picture}}

\newsavebox{\osquareb}
\savebox{\osquareb}(2.3,2){
\begin{picture}(1,1)(1,-1.25)
 \thinlines
 \put(-.5,-.5){\framebox(.5,.5){}}
\end{picture}}

\newsavebox{\squareb}
\savebox{\squareb}(2.3,2){
\begin{picture}(1,1)(1,-1.25)
 \thinlines
 \put(-.75,-1){\text{\rule{1.0pt}{1.0pt}}}
\end{picture}}

\newcommand{\preset}[2]{{}^{#1}{}#2}

\newcommand{\simil}[2]{\langle #1 \| #2 \rangle}

\newcommand{\ideal}[1]{\alg #1}

\newcommand{\structure}[1]{\mathfrak{#1}}

\newcommand{\theory}[1]{\mathsf {#1}}

\newcommand{\con}{{\operatorname{Con}}}
\newcommand{\Con}{{\operatorname{Con}}}
\newcommand{\lang}[1]{{\mathcal {#1}}}

\newcommand{\geom}[1]{{\calcmd{#1}}}
\newcommand{\system}[1]{\mathfrak{#1}}

\newcommand{\proves}{\vdash}
\newcommand{\forces}{\Vdash}

\newcommand{\andmath}{\;\mbox{\rm and}\;}
\newcommand{\ormath}{\;\mbox{\rm or}\;}
\newcommand{\powerset}{{\operatorname{\mathcal P}}}

\newcommand{\proj}[1]{\breve{#1}}
\newcommand{\state}{{\boldsymbol\sigma}}

\newcommand{\bs}[1]{{\boldsymbol{#1}}}

\newcommand{\im}{\operatorname{im}}
\newcommand{\dom}{\operatorname{dom}}
\newcommand{\lspan}{\operatorname{span}}
\newcommand{\Tr}{\operatorname{Tr}}
\newcommand{\ip}[2]{\langle #1 | #2 \rangle}
\newcommand{\bip}[2]{\bigl\langle #1 \big| #2 \bigr\rangle}
\newcommand{\biglip}[2]{\left<\left. #1 \,\right|\, #2 \right>}
\newcommand{\bigrip}[2]{\left< #1 \,\left|\, #2 \right.\right>}

\newlength{\tinysupscrlen}
\newlength{\tinysubscrlen}
\newsavebox{\vrefbox}

\newcommand{\id}{\boldsymbol1}

\newcommand{\chapnotepageref}[1]%
  {{\sc[Refer to p.~\pageref{#1}.]}%
  \vspace{0.5\baselineskip}}

\newcommand{\direction}[1]{\cnum{#1}}

\renewcommand{\Huge}{\huge}

\newcommand{\bigtimes}[2]%
 {{\text{\raisebox{-.6ex}%
 {$\underset{#1}{\overset{#2}{\mbox{\Huge$\times$}}}$}}}}

\renewcommand{\thepremprime}{\arabic{premprime}${}'$}

\renewcommand{\thepremprimestar}{\arabic{premprimestar}${}'{}^*$}

\renewcommand{\thepremdprime}{\arabic{premdprime}${}''$}

\newtheorem{theorem}{Theorem}

\newtheorem{statement}{}
\newcommand{\Element}{\operatorname{Element}}
\newcommand{\Set}{\operatorname{Set}}
\newcommand{\Class}{\operatorname{Class}}
\renewcommand{\opneg}{{\boldsymbol{\compl}}}
\renewcommand{\disptitle}[1]{{\sc #1}}
\newcommand{\entails}{{\Vvdash}}
\newcommand{\nentails}{\not{\hspace{-1ex}\Vvdash}}
\newcommand{\bcompl}{{\pmb{\compl}}}
\newcommand{\bvee}{{\pmb\vee}}
\newcommand{\bwedge}{{\pmb\wedge}}
\newcommand{\bbigvee}{{\pmb\bigvee}}
\newcommand{\bbigwedge}{{\pmb\bigwedge}}
\newcommand{\bimplies}{{\pmb\lcimplies}}
\newcommand{\biff}{{\pmb\lciff}}
\newcommand{\subexp}[1]{{\hat{#1}}}
\newcommand{\parper}{\square}
\newcommand{\semifil}[1]{{[#1]}}

\newcommand{\m}[1]{}
\newcommand{\n}[1]{}
\newcommand{\erf}[1]{}
\newcommand{\prf}[1]{}
\newcommand{\qedex}[1]{\qed}

\begin{document}

\begin{abstract}
There is a natural equivalence relation on representations of the states of a given quantum system in a Hilbert space, two representations being equivalent iff they are related by a unitary transformation.  There are two equivalence classes, with members of opposite classes being related by a conjugate-unitary (anti-unitary) transformation.  These two \emph{conjugacy} classes are related in much the same way as are the two imaginary units of a complex field, and there is \emph{a priori} no basis on which to prefer one over the other in any individual case.  This is potentially problematic in that the choice of conjugacy class of a representation determines the sign of energy and other quantities defined as generators of continuous symmetries of the system in question, so that it would appear that principles like conservation of energy for a compound system may hold or fail depending on relative choices of conjugacy class of representations of its subsystems.  We show that for any finite set of quantum systems there are exactly two ways of choosing conjugacy classes of representations consistent with the usual tensor-product construction for representing the compound system composed of these.  Each is obtained from the other by reversing the conjugacy of all the representations at once.  The relation of unitary equivalence for representations of a single system is therefore uniquely extendible to representations of all systems that can interact with it.
\end{abstract}
\title{Compatibility of representations of quantum systems}
\author{Robert A. Van Wesep}
\pacs{03.65.Ca,03.65.Ta}
\keywords{quantum mechanics, projective representation, conjugate representation, compatible representation, tensor product}
\maketitle

\section{Introduction}

To render a body of physics in the language of mathematics a number of more or less arbitrary conventions must typically be established and observed:  $2\pi$ or not $2\pi$, that is often the question; and the placement of factors of $-1$ is frequently a matter of choice.  The difficulty of maintaining these conventions increases with the indifference of the alternatives and attains its maximum in the case of $\pm i$.  The reason for this is obvious: there is no attribute that distinguishes $i$ from $-i$.  If one were to replace every occurrence of `$i$' in a given text by `$-i$', the validity---indeed, the meaning---of the text would remain unaltered.

Typically, only real quantities have intrinsic physical significance.  Complex numbers sometimes arise merely as a convenience---as, for example, in the analysis of periodic phenomena, where they allow imaginary exponentials to substitute for sine and cosine---and a casual approach to the choice of imaginary unit in these instances is innocuous.  In the quantum theory, however, this choice seems to have greater significance.  For example, the formula
\begin{equation}
\label{eq -i}
U(t) = e^{i H t}
\end{equation}
---or is it
\begin{equation}
\label{eq i}
U(t) = e^{-i H t}\text?
\end{equation}
---defining the hamiltonian $H$ from the time-evolution operators $U(t)$, is fundamental in quantum mechanics.  The choice of sign in the exponent determines the sign of $H$ in this definition, and therefore the sign of the energy, which $H$ measures.

This problem cannot be defined away simply by declaring (\ref{eq -i}) or (\ref{eq i}) as the correct definition.  To see why, suppose $\system S$ is a quantum system and $\vecsp V$ is a Hilbert space suitable for representing the states of $\system S$.  As we will show, there are two classes of representations of $\system S$ in $\vecsp V$, such that any two representations in the same class are unitarily equivalent, while representations in different classes are related by a conjugate-unitary (anti-unitary) operator.  There is no reason \emph{a priori} to prefer either class of representations; there is indeed no attribute in terms of which such a preference could be expressed in general; but the choice has consequences.

The time-evolution operations on $\system S$ are real physical relationships, independent of, and definable without reference to, any mathematical representation.  Once a representation is chosen they are representable as a unitary operator-valued function of a time coordinate, $t \mapsto U(t)$.  From this a hamiltonian is defined---either by (\ref{eq -i}) or (\ref{eq i}), as we will have decided ahead of time.  The hamiltonian, which is an operator on $\vecsp V$, is now rendered physically meaningful via the same representation we started with, and we obtain the energy of $\system S$.  The problem is that the sign of the energy depends on the conjugacy class to which the representation belongs.

This ambiguity may be localized somewhat differently.  Numbers in general, and complex numbers in particular, are not fixed objects, even in the abstract sense acceptable in mathematics.  The phrase `complex numbers' refers to any structure consisting of operations (called `addition' and `multiplication') on an arbitrary set of objects satisfying certain rules, by virtue of which it is a \emph{complex field}.  All complex fields have the same form (they are isomorphic), and any is as good as any other for any purpose.  A Hilbert space $\vecsp V$, properly considered, incorporates a complex field $\mathbb C$ as part of its structure.  Keeping this in mind, let us reconsider the procedure outlined above for coming up with the energy of $\system S$.  As before, there are two equivalence classes of representations of the states of $\system S$ by vectors of $\vecsp V$.  A representation having been chosen, physical time-evolution is represented by a unitary group $t \mapsto U(t)$ as before, and we would now like to define the hamiltonian operator by whichever of (\ref{eq -i}) or (\ref{eq i}) we have decided ahead of time to use.  But the complex field $\mathbb C$ has two imaginary units (square roots of $-1$), either of which may be called `$i$', and the sign of the energy depends on this perfectly arbitrary choice.  In any instance, swapping imaginary units is equivalent to swapping conjugacy classes of representations, so this is the same problem in a different light.

The issue we have raised here for energy arises for any quantity defined as the hermitian generator of a continuous symmetry group, e.g., momentum as the generator of space-translation.  It should be noted, however, that for a given system $\system S$ and Hilbert space $\vecsp V$, once the choice of conjugacy class and imaginary unit is made, it applies to all such relationships.  In practice the choice for a given system or class of systems is made according to \emph{ad hoc} considerations, chief among which is the principle that if the range of energies available to a system is unbounded, it is only unbounded above, never below.  There is nothing that prevents us from taking the opposite position---that energy can only be unbounded below---and we would then make the opposite choice of representation class (or imaginary unit) in each such case.  The kinetic energy of a particle would be $-\tfrac12mv^2$, and putting a flame under a pot of water would draw energy out of it, actually lowering its temperature---which is not unreasonable, since temperature, which is the derivative of energy with respect to entropy, would be negative for most systems, including pots of water, so lowering the temperature increases its magnitude.

Clearly, in general, it would not do to effect such a reversal for one system and not for another.  For example, if the systems can exchange energy then conservation of energy cannot hold for the combined system both before and after the reversal.  It is worth noting that if energy were conserved in the latter case, all hell would break loose, rather literally, as energy poured endlessly from the negative-temperature to the positive-temperature system, increasing the entropy of both without limit and driving both temperatures to infinity.  (Energy would flow from negative to positive temperature because that is the direction of increasing entropy for both systems.  The apparent paradox of energy flowing from low to high temperature is resolved if we realize that negative temperatures are actually not very cold, but rather very hot; they can only be attained by heating past $\infty$---not by cooling past 0, which is impossible.  The more appropriate parameter to describe this is $\beta = 1/T$, the derivative of entropy with respect to energy, for which 0 is a perfectly acceptable value.  For example, in a pure up-down charged spin system on a square grid, whose energy is the sum of the magnetic interactions of adjacent sites, the states of least energy are those with all spins up or all spins down, and the entropy is minimized.  As energy is added, spins flip, and entropy increases until it reaches a maximum, then decreases until a state of maximum energy is attained, in which the spins are arranged in strict alternation, and entropy is again minimized.  At the entropy maximum, $\beta$ is by definition 0, and $T = \pm\infty$.  By definition, entropy increases as energy flows from lower to higher $\beta$, regardless of sign.)

If the systems did not interact, of course, we would be spared this catastrophe; and upon reflection, we see that in this case nothing prevents us from switching the representation class for one system and not the other.  There would be no adverse consequences for the physics of the composite system because they cannot be combined in any significant way.  Since non-interacting systems cannot exchange energy, we would actually even retain energy conservation.

It would appear, then, that if there is an underlying principle that dictates the relative choice of representations (or imaginary units) of diverse systems, its source must lie in the quantum mechanics of composite systems.

To understand clearly the nature of the problem and its solution, we must examine carefully how complex numbers enter into the quantum theory, and what it means to choose, or not to choose, one imaginary unit over the other.  This is obviously a rather delicate matter, and we must be quite precise regarding the mathematical structures with which we deal.  Section~\ref{sec math struc} is devoted to developing an appropriate terminology and some basic principles, culminating with the crucial theorem of Wigner relating projective isometries to linear isometries in separable complex inner-product (SCIP) spaces.  Section~\ref{sec phys struc} relates these to the \emph{similarity structure} of quantum systems, the \emph{similarity} of states $\Psi$ and $\Phi$ being the probability that the answer to the question `is the state $\Psi$?' is `yes' when the state is $\Phi$ (and \emph{vice versa}).

Section~\ref{sec con sym} describes how formulas such as (\ref{eq -i}) and (\ref{eq i}) arise and how the choice of representation affects the sign.  Section~\ref{sec sign energy} shows that this matter of signs is not entirely straightforward:  the conventional choice of sign of energy is in a reasonable sense the wrong one.  Section~\ref{sec time rev} applies these ideas to the operation of time-reversal by way of illustration.  These three sections are not essential to the argument and may be omitted.  

In Section~\ref{sec repr comp} we state and prove the main theorem, which we use in Section~\ref{sec comp repr} to define the notion of \emph{compatibility} of representations of distinct physical systems in terms of their similarity structures \visavis that of that of the compound system formed from them.  The familiar tensor product representation of a compound system is applicable iff compatible representations are used for all the constituent subsystems.  Given a representation of any one system, this principle mandates the conjugacy class of the representation of any system with which it may be compounded, quite independently of the dynamics (time-translation behaviors) and any other physical properties of the systems in question.

One would think that a matter so basic to quantum mechanics would have been dealt with early in the history of the theory, and perhaps it has been.  I have been unable to find any reference to the problem in the current literature (i.e., that which is available via the usual databases of scientific literature or Google.  I would appreciate any relevant references the reader might know of. 

\section{Mathematical structures}
\label{sec math struc}

We define a \emph{real field} to be a complete archimedean ordered field.  A real field is therefore a structure
\[
\mathbb R = (R, <, +, \cdot),
\]
where $R$ is a set, $<$ is a binary relation on $R$, and $+$ and $\cdot$ are binary operations on $R$, with the several properties listed above, all of which apply to real numbers.  $R$ is the \emph{universe} of the structure.  In general we use `$|\cdot|$' to denote the universe of a structure, so $|\mathbb R| = R$ in this case.  (`$\cdot$' does double duty in this article.  It denotes multiplication, and it is a placeholder for arguments of functions and relations, as in `$|\cdot|$'.)  Any two real fields $\mathbb R = (R, <, +, \cdot)$ and $\mathbb R' = (R', <', +', \cdot')$ are \emph{isomorphic}, i.e., there is a bijection (a one-one correspondence) $\iota: R \to R'$ such that for all $x,y \in R$,
\begin{eqnarray*}
x < y &\iff& \iota(x) <' \iota(y)\\
\iota(x+y) &=& \iota(x) +' \iota(y)\\
\iota(x\cdot y) &=& \iota(x) \cdot' \iota(y).
\end{eqnarray*}
Hence any real field serves as well as any other as ``the real numbers'', and it is customary to use the concept of real number without specifying any particular real field.

Similarly, a \emph{complex field} is a structure
\[
\mathbb C = (C, R, <, +, \cdot)
\]
such that $+$ and $\cdot$ are binary operations on $C$, $R \subseteq C$, $<$ is a binary relation on $R$, $(C, +, \cdot)$ is a field, $(R, <, +, \cdot)$ is a real field, and there exists $j \in C$ such that $j \cdot j = -1$ and for all $z \in C$ there exist $x, y \in R$ such that $z = x + y \cdot j$.  Any two complex fields are isomorphic.  We refer to an element $j$ such that $j \cdot j = -1$ as an \emph{imaginary unit}.  Clearly, any complex field has two imaginary units, each the additive inverse of the other.  For a complex field $\mathbb C = (C, R, <, +, \cdot)$ we define $|\mathbb C|$ to be $C$.

We have noted above that given any two real fields there is an isomorphism that relates them. It is easily shown that this isomorphism is unique.  Given two complex fields, on the other hand, there are two isomorphisms that relate them.  These are easily constructed.  Given complex fields $\mathbb C = (C, R, <, +, \cdot)$ and $\mathbb C' = (C', R', <', +', \cdot')$, their real parts are related by a unique isomorphism, say $\rho : R \to R'$.  Let $j \in R$ and $j' \in R'$ be such that $j \cdot j = -1$ and $j' \cdot' j' = -'1'$.  Any isomorphism $\iota$ of $\mathbb C$ with $\mathbb C'$ must extend $\rho$ and must take each imaginary unit of $\mathbb C$ to an imaginary unit of $\mathbb C'$.  So either $\iota(j) = j'$ or $\iota(j) = -' j'$.  It is easily shown that each of these choices extends uniquely to an isomorphism of $\mathbb C$ with $\mathbb C'$.

A \emph{complex field with distinguished imaginary unit}, or \emph{complex field wdiu}, is a structure $\mathbb C = (C, R, <, +, \cdot, j)$, where $j$ is an imaginary unit for $(C, R, <, +, \cdot)$.  An isomorphism of two complex fields wdiu must by definition take the distinguished imaginary unit of one to that of the other, so in this case the isomorphism is unique.

The closest we can come to the notion of a real number \emph{in abstracto} is a function $F$ that assigns to each real field $\mathbb R$ an element $F(\mathbb R) \in |\mathbb R|$ in such a way that for any two real fields $\mathbb R$ and $\mathbb R'$, $F(\mathbb R)$ and $F(\mathbb R')$ correspond under the (unique) isomorphism of $\mathbb R$ with $\mathbb R'$.  This does not work for complex fields, as the isomorphisms are not unique, but it does work for complex fields wdiu. We may regard $i$ as a specific complex number only in the weak sense that it is the distinguished imaginary unit of any complex field wdiu.

An \emph{automorphism} of a structure $\structure S$ is an isomorphism of $\structure S$ with itself.  The \emph{automorphism group} of $\structure S$ is the set of automorphisms of $\structure S$, together with the composition operation.  A real field has only the identity automorphism, while a complex field has also the conjugation operation.  A complex field wdiu has only the identity automorphism.

A complex vector space is properly regarded as a structure $\vecsp V = (V, C, R, +_v, \cdot_s, <, +, \cdot)$, where $+_v : V \times V \to V$ and $\cdot_s : C \times V \to V$ are vector addition and scalar multiplication, and $(C, R, <, +, \cdot)$ is a complex field.  If $j \in C$ is such that $j\cdot j = -1$, then $\vecsp V = (V, C, R, +_v, \cdot_s, \ip{\cdot}{\cdot}, <, +, \cdot, j)$ is a complex vector space with distinguished imaginary unit or wdiu.  If we add a positive definite hermitian product $\ip{\cdot}{\cdot} : V \times V \to C$ we have an inner-product space, with or without a distinguished imaginary unit.  An inner-product space is separable iff every set of pairwise orthogonal vectors is countable.  We use `SCIP' to abbreviate `separable complex inner-product'.  A Hilbert space is a SCIP space that is \emph{complete}, i.e., every Cauchy sequence has a limit, or, equivalently, if $\sum_{n=1}^\infty \|u_n\|^2$ exists, then $\sum_{n=1}^\infty u_n$ exists.  We will use `$V$', `$C$', etc., and their congeners---e.g., $V'$,$V_0$---to denote the corresponding components of spaces denoted by $\vecsp V$ and its respective congeners.

Any SCIP space may be embedded as a dense subspace of a Hilbert space (by adjoining limits of all Cauchy sequences).   An isomorphism of a dense subspace of a Hilbert space with a dense subspace of another Hilbert space extends uniquely to an isomorphism of the complete spaces, so the extension of a SCIP space to a Hilbert space is essentially unique, and much of the discussion of SCIP spaces is more conveniently carried out in terms of the corresponding Hilbert spaces.  This is customary in quantum mechanics, but we will nevertheless be concerned primarily with SCIP spaces, as these are most closely tied to the physical structure of quantum mechanical systems.

An isomorphism of SCIP spaces
\[
\vecsp V = (V, C, R, +_v, \cdot_s, \ip{\cdot}{\cdot}, <, +, \cdot)
\]
and
\[
\vecsp V' = (V', C', R', +'_v, \cdot'_s, \ip{\cdot}{\cdot}', <', +', \cdot')
\]
is technically a triple $\langle U, \eta, \rho \rangle$, where $U : V \to V'$, $\eta : C \to C'$, $\rho : R \to R'$, and the obvious conditions are satisfied.  In particular, $\rho$ is an isomorphism of the respective real fields, and $\eta$ is an isomorphism of the respective complex fields.  Since $\rho$ is uniquely determined, we need not mention it, so we will indicate an isomorphism of SCIP spaces as $\langle U, \eta \rangle$.

If $(C, +, \cdot) = (C', +', \cdot')$, i.e., if $\vecsp V$ and $\vecsp V'$ incorporate the same complex field, then $\eta$ is either the identity or the conjugation map.  In the former case $U$ is a unitary map, in the latter case it is \emph{conjugate unitary}, also called \emph{anti-unitary}.  In either case, we say that $U$ is \emph{$\eta$-unitary}; if we do not wish to specify $\eta$, we say $U$ is \emph{$\star$-unitary}.  If $\vecsp V$ and $\vecsp V'$ are SCIP spaces with distinguished imaginary units we define \emph{unitary} and \emph{conjugate-unitary} maps from $V$ to $V'$ in the obvious way.  

Given a SCIP space $\vecsp V = (V, C, R, +_v, \cdot_s, \ip{\cdot}{\cdot}, <, +, \cdot)$, we form its \emph{projective space} $\breve{\vecsp V} = (\breve V, R, \simil{\cdot}{\cdot}, <, +, \cdot)$, where
\begin{enumerate}
\item $\breve V$ is the set of \emph{rays}, i.e., 1-dimensional subspaces, of $V$; and
\item $\simil{\cdot}{\cdot} : \breve V \times \breve V \to R$ is the \emph{similarity} operation defined by
\[
\simil{r_0}{r_1} = |\ip{v_0}{v_1}|^2,
\]
where $v_0 \in r_0$ and $v_1 \in r_1$ are arbitrary vectors with norm 1.
\end{enumerate}
We define the \emph{similarity} of two nonzero vectors and the similarity of a nonzero vector and a ray by replacing each vector by the ray that contains it.

We define $\breve v$ to be the ray containing $v$ for any nonzero $v \in V$.  Note that there is no notion of addition or multiplication of rays, and the range of the similarity operation is the set $R$ of real numbers, so the complex numbers of $C$ play no part in the structure of a projective space.  For the purpose of this article we define a \emph{similarity space} to be a structure $(S, R, \simil{\cdot}{\cdot}, <, +, \cdot)$ that is isomorphic to the projective space of a SCIP space. 

Clearly any isomorphism $\langle U, \eta \rangle$ of one SCIP space $(V, C, R, +_v, \cdot_s, \ip{\cdot}{\cdot}, <, +, \cdot)$ with another $(V', C', R', +'_v, \cdot'_s, \ip{\cdot}{\cdot}', <', +', \cdot')$ induces an isomorphism $\breve U$ of their projective spaces $(\breve V, R, \simil{\cdot}{\cdot}, <, +, \cdot)$ and $(\breve V', R', \simil{\cdot}{\cdot}', <', +', \cdot')$ by the prescription that for any $v \in V$, $\breve U \breve v = (Uv)\spbreve$.  The following theorem of Wigner\cite{Wigner:1931} says that all isomorphisms of similarity spaces are obtainable in this way.  (Wigner's statement and proof were specific to the unitary, as opposed to conjugate-unitary, case.  See \cite[Chapter 2, Appendix A]{Weinberg:1995} for a general proof.)
\begin{theorem}
\label{thm Wigner}
Suppose $\vecsp V$ and $\vecsp V'$ are SCIP spaces, and suppose $\iota : \breve V \to \breve V'$ is an isomorphism of their projective spaces.  Then there exists an isomorphism $\langle U, \eta \rangle$ of $\vecsp V$ with $\vecsp V'$, such that for all $v \in V$, $(Uv)\proj{\,\,} = \iota(\proj v)$.
\end{theorem}
In particular, any automorphism of the projective space $\proj{\vecsp V}$ is induced by an automorphism of the vector space $\vecsp V$.

It is easy to show that if $\dim V \ge 2$, $A$ and $A'$ are $\star$-linear operators on $\vecsp V$, and for all $v \in V$, $Av$ and $A'v$ are proportional, then $A$ and $A'$ are proportional.  In particular, if $\langle U, \eta \rangle$ and $\langle U', \eta' \rangle$ are automorphisms of $\vecsp V$, then $\breve U = \breve U'$ iff for some $\alpha \in C$ of norm 1 (a \emph{phase factor}), $U = \alpha U'$.  To avoid irrelevant complications, we will assume henceforth that all our vector spaces have dimension at least 2. It follows that a given automorphism $\mu$ of $\proj{\vecsp V}$ is induced either by a unitary or by a conjugate-unitary operator, but not both, and we say that $\mu$ is \emph{unitary} or \emph{conjugate-unitary} on this basis.

Now suppose $\mu$ is an automorphism of $\proj{\vecsp V}$ and $\iota$ is an isomorphism of $\proj{\vecsp V}$ with another projective space $\proj{\vecsp V}'$.  Then $\mu' = \iota \circ \mu \circ \iota^{-1}$ is an automorphism of $\proj{\vecsp V}'$.  By Wigner's theorem there is an isomorphism $\langle W, \theta \rangle$ of the underlying vector spaces $\vecsp V$ and $\vecsp V'$ that induces $\iota$.  If $\mu$ is represented by $\langle U, \eta \rangle$ then $\mu'$ is represented by $\langle W, \theta \rangle \circ \langle U, \eta \rangle \circ \langle W, \theta \rangle^{-1} = \langle W \circ U \circ W^{-1}, \theta \circ \eta \circ \theta^{-1} \rangle$.  The automorphism $\theta \circ \eta \circ \theta^{-1}$ of $(C', R', <', +', \cdot')$ has the same conjugacy-type (identity or conjugation map) as $\eta$, so $\mu'$ has the same conjugacy type as $\mu$.

Now suppose $\structure S = (S, R, \simil{\cdot}{\cdot}, +, \cdot)$ is a similarity space and $\mu$ is an automorphism of $\structure S$.  Let $\vecsp V = (V, C, R, \ip{\cdot}{\cdot}, <, +, \cdot)$ be a SCIP space such that $\structure S$ is isomorphic to $\proj{\vecsp V}$, say by an isomorphism $\zeta$.  Then $\nu = \zeta \circ \mu \circ \zeta^{-1}$ is an automorphism of $\proj{\vecsp V}$.  Let $\zeta'$ be another isomorphism of $\structure S$ with a projective space $\proj{\vecsp V}'$, and let $\nu' = \zeta' \circ \mu \circ \zeta'{}^{-1}$ be the corresponding automorphism of $\proj{\vecsp V}'$.  Then $\iota = \zeta' \circ \mu \circ \zeta^{-1}$ is an isomorphism of $\proj{\vecsp V}$ with $\proj{\vecsp V}'$, and $\nu' = \iota \circ \nu \circ \iota^{-1}$, so $\nu$ and $\nu'$ have the same conjugacy-type, as noted in the preceding paragraph.  We define the conjugacy-type of $\mu$ to be the common conjugacy-type of all of its representations by projective automorphisms, i.e., an automorphism of a similarity space is unitary or conjugate-unitary according as its representations by projective automorphisms are unitary or conjugate-unitary.

It is perhaps a little surprising that the abstract structure of a similarity space, in which complex numbers have no direct role, should contain within it this element of conjugacy.  Note that while isomorphisms of a similarity space $\structure S$ with a different similarity space $\structure S'$ do not individually have any attribute of conjugacy, they nonetheless fall into two equivalence classes, where $\iota$ and $\iota'$ are equivalent iff $\iota' \circ \iota^{-1}$ (equivalently, $\iota \circ \iota'{}^{-1}$) is unitary.

Lest the reader to whom conjugate-linearity is a novel concept suppose that it is of no physical relevance and might safely be ignored, we mention that there is at least one operation in quantum mechanics that can only be represented by a conjugate-unitary operator, viz., time-reversal, as we show in Section~\ref{sec con sym}.

\section{Physical structure}
\label{sec phys struc}

Suppose $\system S$ is a physical system.  We suppose that for each state $\boldsymbol\sigma$ of $\system S$ there is a measurement that corresponds to the question `is $\system S$ in the state $\boldsymbol\sigma$?', and we call this measurement $M_{\boldsymbol\sigma}$.  It is a fundamental principle of quantum mechanics that for any states $\boldsymbol\sigma_0$ and $\boldsymbol\sigma_1$, the probability that $M_{\boldsymbol\sigma_0}$ yields a positive result when the state is $\boldsymbol\sigma_1$ is equal to the probability that $M_{\boldsymbol\sigma_1}$ yields a positive result when the state is $\boldsymbol\sigma_0$.  We define the \emph{similarity} $\simil{\boldsymbol \sigma _0}{\boldsymbol\sigma_1}$ of states $\boldsymbol\sigma_0$ and $\boldsymbol\sigma_1$ to be this probability.  We define the \emph{(physical) structure} of $\system S$ to be $(\boldsymbol S, R, \simil{\cdot}{\cdot}, <, +, \cdot)$, where $\boldsymbol S$ is the set of physical states, $(R, <, +, \cdot)$ is a real field, and $\simil{\cdot}{\cdot}$ is the physical similarity operation with values in $R$.  Of course, this is really \emph{a} physical structure of $\system S$, since the real field is arbitrary; but given the uniqueness of isomorphisms of real fields, all of these are related by unique isomorphisms that are the identity on $\boldsymbol S$.  Note that our attitude here is that states $\bs \sigma \in \bs S$ are actual \emph{physical} entities or attributes, intrinsic to $\system S$, not mathematical abstractions.

The \emph{superposability principle} of quantum mechanics, properly formulated, implies that $\boldsymbol S$ is the union $\bigcup_n \boldsymbol S_n$ of countably many sets, called \emph{superposability} or \emph{superselection sectors}, such that for each $n$, $(\boldsymbol S_n, R, \simil{\cdot}{\cdot}, <, +, \cdot)$ is a similarity space, and for all $m \neq n$, $\boldsymbol u \in \boldsymbol S_m$, and $\boldsymbol v \in \boldsymbol S_n$, $\simil{\boldsymbol u}{\boldsymbol v} = 0$.  Each superposability sector comprises all states superposable with a given state, and states in distinct superposability sectors are orthogonal.  To avoid unnecessary complications, we will assume that there is only one superposability sector for each of the systems under consideration---equivalently, we concern ourselves with a single superposability sector of each system.  Hence $\system S = (\boldsymbol S, R, \simil{\cdot}{\cdot}, <, +, \cdot)$ is isomorphic to the projective space of a SCIP space $\vecsp V$.  Note that we do not require that $\vecsp V$ be complete, i.e., that it be a Hilbert space.  This is because \emph{every} nonzero vector of $\vecsp V$ is supposed to represent a physical state.  If $\bar{\vecsp V}$ is the canonical extension of $\vecsp V$ to a Hilbert space, and $A$ is a selfadjoint operator on $\bar{\vecsp V}$ representing an unbounded quantity---e.g., energy in most cases of interest---there are nonzero vectors in $\bar{\vecsp V}$ that are not in the domain of $A$.  For these the energy is not defined (it is typically ``infinite''), and they cannot represent physical states.

As noted in the preceding section for similarity spaces in general, by Wigner's theorem, given any isomorphism $\iota$ of $\structure S$ with a projective space $\proj{\vecsp V}$ derived from a SCIP space $\vecsp V$, any automorphism $\mu$ of $\system S$ is represented by an automorphism $\langle U, \eta \rangle$ of $\vecsp V$ in the sense that $\mu = \iota^{-1} \circ \proj U \circ \iota$.  Assuming as we do that $\dim V \ge 2$, $U$ is uniquely determined by $\iota$ and $\vecsp V$ up to a phase factor, and the conjugacy-type of $\eta$ (either the identity or the conjugation map) is uniquely determined by $\mu$, independent of representation.

\section{Continuous symmetry groups and their generators}
\label{sec con sym}

The purpose of this section is to demonstrate the effect of the choice of representation on observables that arise as generators of continuous symmetry groups.  It is largely motivational and is not required for the main result of this article. 

The physics of a system $\structure S$ is mostly a matter of its behavior under the action of symmetry groups.  In particular, the dynamic of $\structure S$ is just its behavior under time-translation.  Given any symmetry operation on $\structure S$ and any procedure by which two states $\boldsymbol \sigma$ and $\boldsymbol \sigma'$ might be distinguished, there is an entirely equivalent procedure that involves first transforming the states by the symmetry operation, and then applying the given procedure.  It follows that the action of a symmetry operation on $\structure S$ is an automorphism of $(\boldsymbol S, \mathbb R, \simil{\cdot}{\cdot})$.  (Note that we have indicated the strictly numerical features of $\structure S$, viz., $R$, $<$, $+$, and $\cdot$, by the single symbol `$\mathbb R$'.  We may also use `$\mathbb R$' loosely to refer to $R$.  We will use this notation and the corresponding notation for complex fields without further comment.) The composition of any two symmetry operations is again a symmetry operation, and any symmetry operation has an inverse, so the symmetries of a given system from a group under composition.  Accordingly, the theory of groups of automorphisms of similarity spaces is central to the quantum theory.

As the reader is well aware, even though the physical structure of $\structure S$ and its automorphisms, in which complex numbers are nowhere to be seen, constitute physical reality, it is the representations of these entities in complex linear spaces that get all the mathematical attention in quantum mechanics.  In the preceding section we have described the representation of similarity spaces---and therefore of physical systems---in linear spaces.  We now turn to the topic of groups of automorphisms of similarity spaces.  We will restrict our remarks to so-called 1-parameter groups $\tau \mapsto A_\tau$, where $\tau$ ranges over $\mathbb R$, and $A_{\tau + \tau'} = A_\tau \circ A_{\tau'}$.  

As noted above, a physical system $\system S = (\boldsymbol S, \mathbb R, \ip{\cdot}{\cdot})$ is isomorphic to a projective space, say $\proj{\vecsp V} = (\proj V, \mathbb R, \simil{\cdot}{\cdot})$, so the theory of automorphisms of $\system S = (\boldsymbol S, \mathbb R, \ip{\cdot}{\cdot})$ is just that of $\proj{\vecsp V}$.  Suppose $\tau \mapsto A_\tau$ is a 1-parameter group of automorphisms of $\system S$ and $\iota : \bs S \to \proj V$ is an isomorphism.  Let
\begin{equation}
P_\tau = \iota \circ A_\tau \circ \iota^{-1}.
\end{equation}
$\tau \mapsto P_\tau$ is a 1-parameter group of automorphisms of $\proj{\vecsp V}$.  We know from Wigner's theorem that for each $\tau \in \mathbb R$, there is an automorphism $\langle U, \eta \rangle$ of $\vecsp V = (V, \mathbb C, \ip{\cdot}{\cdot})$ such that
\[
\proj U = P_\tau,
\]
and that $\eta$ is uniquely determined and $U$ is determined up to a phase factor by this condition.  We refer to choosing a representative $U_\tau$ as ``adjusting the phase''.  Following the usual policy in physics of assuming as much regularity, or smoothness, of functions as is needed to allow the mathematical analysis to proceed, we suppose that the map $\tau \mapsto P_\tau$ is sufficiently smooth that $\eta$ is a continuous function of $\tau$.  Since $\eta(0) = \id$, the identity automorphism of $\mathbb C$, the same is true of $\eta$ for all $\tau$, and we will omit its mention in the remainder of this discussion.

We note in passing that the requirement of unitarity by virtue of continuity does not apply to discrete symmetries, such as space-inversion, time-reversal, and charge-conjugation.  In particular, as mentioned above and proven below, time-reversal is necessarily represented by a conjugate-unitary operator.

To obtain the usual mathematical setting of quantum mechanics, we require that the phases of the $U_\tau$s be adjusted in such a way that the map $\tau \mapsto U_\tau$ is itself a group.  (See the appendix for a discussion of this.)  We require also that $\tau \mapsto U(\tau)$ be sufficiently smooth that the derivative
\[
{dU_\tau u \over d\tau},
\]
for any $\tau \in \mathbb R$, exist for a sufficiently large set of vectors $u \in \vecsp V$.  Clearly this derivative is a linear function of $u$ and its domain is a subspace of $\vecsp V$.  Define an operator $K$ by
\[
K u = \left. {d U_\tau u \over d\tau} \right|_{\tau = 0}.
\]
Since $U_\tau$ is unitary for all $\tau$,
\[
0
=
\left.{d \ip{U_\tau u}{U_\tau v}\over d\tau}\right|_{\tau = 0}
=
\ip{Ku}{v} + \ip{u}{Kv},
\]
so $K$ is skew-hermitian.

Conversely, if $K$ is skew-hermitian, and we let $W_\tau = \exp(\tau K)$, then for every $u, v \in \vecsp V$ and $\tau \in \mathbb R$,
\[
{d \ip{W_\tau u}{W_\tau v}\over d\tau}
 =
\ip{K W_\tau u}{W_\tau v} + \ip{W_\tau u}{K W_\tau v}
 =
0,
\]
i.e., $\ip{W_\tau u}{W_\tau v}$ is a constant function of $\tau$.  Since $W_0 = \id$, $\ip{W_\tau u}{W_\tau v} = \ip{u}{v}$ for all $u, v$, and $\tau$, i.e., $W_\tau$ is unitary for all $\tau$.  Since $\exp (\tau K + \tau' K) = \exp (\tau K) \exp (\tau' K)$, $\tau \mapsto W_\tau$ is a 1-parameter unitary group, and indeed, $W_\tau = U_\tau$.  In other words, any sufficiently smooth 1-parameter unitary group is of the form $\tau \mapsto \exp(\tau K)$ for a skew-hermitian operator $K$.  (This heuristic discussion is properly framed in terms of selfadjoint operators on Hilbert space, but as this discussion is for illustrative purposes only, we make no attempt at rigor.  See the appendix for more on this.)

Recall that $U_\tau$ is defined by $P_\tau$ only up to a phase factor.  Suppose $c \in \mathbb C$ is imaginary, and let $U'_\tau = \exp(\tau c) U_\tau$.  Then $U'_\cdot$ is another unitary group that represents $P_\cdot$.  The generator of this group is $K' = K + c \id$.  All smooth unitary representations of $P_\cdot$ are of this form.  Hence the expectation and the eigenvalues just mentioned are determined by $P_\cdot$ up to an additive imaginary constant.  In the example of the time-translation group this ambiguity corresponds to the fact that, in non-relativistic physics, absolute energy is meaningless---physical laws deal only in differences of energy.  As mentioned in the introduction, this is not true in relativistic physics, and the quantum mechanical reason for this is that time-translation is incorporated into the larger and nonabelian Poincar\'e group.  In this connection the theorem of Bargmann\cite{Bargmann:1954} is relevant.  See also \cite[p.~234]{Bogolubov:1975} for a general discussion of Bargmann's and Wigner's theorems and related issues.

There is, however, a deeper and unavoidable ambiguity in the choice of generators, which arises from the choice of conjugacy class of the initial projective representation $\iota : \bs S \to \proj V$ and applies to all continuous symmetries.  Recall that the projective group $\tau \mapsto P_\tau$ is a mathematical representation of the physical group, $\tau \mapsto A_\tau$.  Specifically, $P_\tau = \iota \circ A_\tau \circ \iota^{-1}$, where $\iota : \boldsymbol S \to \proj V$ is the particular isomorphism by which physical states are represented by rays in $\vecsp V$.  Any two such isomorphisms are related by an automorphism $\mu$ of $\proj{\vecsp V}$, which is represented by an automorphism $\langle W, \eta \rangle$ of $\vecsp V = (V, \mathbb C, \mathbb R, \simil{\cdot}{\cdot})$, where $\eta$ is an automorphism of $\mathbb C$ and $W$ is a unitary or conjugate-unitary operator according as $\eta$ is the identity or the conjugation map.

Such a change of representation of physical states in $\proj{\vecsp V}$ leads to a change of representation of the physical symmetry group whereby each group operator $U_\tau$ is replaced by $U'_\tau = W U_\tau W^{-1}$, and the generator $K$ is replaced by $K' = W K W^{-1}$.  If $W$ is unitary (i.e., $\eta$ is the identity automorphism of $\mathbb C$) nothing much has changed.  In particular, the spectrum of $K'$ is that of $K$.  For suppose $u$ is a $\kappa$-eigenvector of $K$ (with the usual work-around for continuous spectra).  Let $u' = W u$.  Then
\[
K' u'
 =
W K W^{-1} (W u) = W \kappa u = \kappa Wu = \kappa u',
\]
so $u'$ is a $\kappa$-eigenvector of $K'$.  If $W$ is conjugate-unitary, however,
\[
K' u'
 =
W \kappa u = -\kappa Wu = -\kappa u',
\]
since $\kappa$ is imaginary.

In a sense, this makes no difference, since to obtain (numerically and physically) real magnitudes for expectations and eigenvalues, we must multiply $K$ by an imaginary unit, and we have two choices within $\mathbb C$ for this unit, say $j$ and $j'$.  $jK$ and $j'K'$ have the same spectrum.   As discussed in the introduction, there is no basis on which to prefer one conjugacy class of representation in a given projective space over the other, and there is no basis on which to prefer one imaginary unit over the other.  The effect of changing either one of these is to multiply the hermitian generators of all continuous symmetries of $\system S$ by $-1$; the effect of changing both at once is to leave things the same.  We will refer to these choices as `setting the sign' of the corresponding physical quantities.  Note that setting the sign of one such quantity sets all the others for a given system.  We take energy as a paradigm.

\section{The sign of energy}
\label{sec sign energy}

As discussed in the introduction, certain conventions have historically been invoked---if only implicitly---to set the sign of energy.  For example, for many systems the  energy is bounded below and unbounded above; this comports with our notion of energy as something of which a system has a definite quantity---more can always be added, but only so much can be removed before it is all gone.  Alternatively---and, as it happens, consistently---the sign of energy is chosen so that entropy increases with increasing energy, i.e., temperature is positive.  (Systems like the spin system mentioned above, which have states of negative temperature, are idealizations arrived at by ignoring physically important modes such as vibration.)

The purpose of this section is to show that familiar conventions regarding the sign of the generators of continuous symmetries are not always consistent.  In particular, energy typically gets the opposite sign to that which would follow from a uniform treatment of time- and space-translation.  This is true even in classical mechanics, where continuous symmetries lead to conserved quantities via the lagrangian formulation in a way entirely analogous to quantum mechanics.  So as not to extend this digression unduly, we will restrict our remarks to the quantum case.

Consider, for example, a non-relativistic free particle in one dimension.  A convenient representation space for the instantaneous state is the space of $L^2$ (i.e., square-integrable) functions $\psi : \mathbb R \to \mathbb C$, with the familiar inner product:  $\ip{\psi}{\phi} = \int_{-\infty}^\infty dx\, \overline{\psi(x)} \phi(x)$.  All integrals will be over $(-\infty,\infty)$, and we will assume sufficiently rapid decrease of all functions of $x$ and their derivatives so that the following computation is valid:
\[
\begin{split}
\int dx\, f(x) { d \over dx} g(x)
 &=
\left. f(x) g(x) \right|_{-\infty}^\infty - \int dx\, {d \over dx} f(x)  g(x)\\
 &=
- \int dx\, {d \over dx} f(x)  g(x).
\end{split}
\]
By this means we show that $j \partial / \partial x$ is hermitian for either imaginary unit $j$.  We let $\partial_x = \partial / \partial x$ and $\partial_t = \partial / \partial t$.

The global state is represented by a function $\Psi : \mathbb R \times \mathbb R \to \mathbb C$ that satisfies a Schr\"odinger equation:
\begin{equation}
\label{eq schro}
j \partial_t \Psi(t,x)
 =
{1 \over 2m} \partial_x^2 \Psi(t,x),
\end{equation}
where $j$ is an imaginary unit.

For each $t \in \mathbb R$, the function $\Psi_t$, defined by $\Psi_t(x) = \Psi(t,x)$, represents the instantaneous state at time $t$.  Let $\vecsp V$ be the space of global statefunctions.  The inner product on $\vecsp V$ is given by
\[
\ip{\Psi}{\Phi} = \ip{\Psi_t}{\Phi_t},
\]
where $t \in \mathbb R$ is arbitrary.  Since $\partial_x^2 = - (i \partial_x)^2$ is hermitian, this is independent of the choice of $t$.  

By definition, an operation $A$ on states of physical systems is a symmetry operation iff for any states $\Psi_1, \dots, \Psi_N$ of systems $\system S_1, \dots, \system S_N$, all relationships are the same for $A\Psi_1, \dots, A\Psi_N$ as for $\Psi_1, \dots, \Psi_N$.  Applying this general rule to observers, which are, after all, physical systems, we have:
\begin{statement}
\label{stmnt standard symm}
For any state $\Psi$ of a system $\system S$ and $\Theta$ of an observer $\system O$, $A\Psi$ appears to $A \Theta$ as $\Psi$ appears to $\Theta$.
\end{statement}
According to this rule, the (unitary) operator $U^\text{x}_\xi$ given by
\[
\{ U^{\text{x}}_\xi \Psi \}(t,x + \xi) = \Psi(t, x)
\]
represents spatial translation by the amount $\xi$.  The skew-hermitian generator of the group $\xi \mapsto U^\text{x}_\xi$ is given by
\[
\begin{split}
\{K^{\text{x}} \Psi\}(t,x)
 &=
{d \over d\xi}\{ U^{\text{x}}_\xi \Psi \}(t,x)
 =
{d \over d\xi}\Psi(t,x-\xi)\\
 &=
- \partial_x \Psi(t,x).
\end{split}
\]
The same rule applied to time-translation gives the group $U^\text{t}_\cdot$ defined by
\[
\{ U^{\text{t}}_\tau \Psi \}(t + \tau, x) = \Psi(t, x),
\]
with the skew-hermitian generator
\begin{equation}
\label{eq gen time-tran}
\begin{split}
\{K^{\text{t}} \Psi\}(t,x)
 &=
- \partial_t \Psi(t,x)
 =
{j \over 2m} \partial_x^2 \Psi(t,x)\\
 &=
{j \over 2m} \{K^{\text{x}}{}^2 \Psi\}(t,x).
\end{split}
\end{equation}
The momentum and energy operators are obtained by multiplying $K^\text{x}$ and $K^\text{t}$ by imaginary units, of which there are two.  To decide which one we want to use, we examine the motion of the particle.

The expectation of the position at time $t$ is
\[
\ip{\Psi_t}{X \Psi_t}
 =
\int dx\, \overline{\Psi(t,x)} x \Psi(t,x),
\]
where $X$ is the position operator on the instantaneous statespace given by
\[
\{X\psi\}(x) = x \psi(x);
\]
and
\begin{widetext}
\[
\begin{split}
{ d \over dt} \ip{\Psi_t}{X\Psi_t}
 &=
\int dx\, \big(\overline{\partial_t \Psi(t,x)} x \Psi(t,x)
 + \overline{\Psi(t,x)} x \partial_t \Psi(t,x) \big)
 =
{1\over 2m} \int dx\, \big(
 j \overline{\partial_x^2 \Psi(t,x)} x \Psi(t,x)
 - j \overline{\Psi(t,x)} x\partial_x^2 \Psi(t,x) \big)\\
 &=
{1 \over 2m} \int dx\, \big(
 j \overline{\Psi(t,x)} \partial_x^2 (x \Psi(t,x))
 - j \overline{\Psi(t,x)} x\partial_x^2 \Psi(t,x) \big)
 =
{j \over m} \big(
\overline{\Psi(t,x)} \partial_x \Psi(t,x) \big)
 =
-{j \over m} \ip{\Psi_t}{ \{K^{\text{x}} \Psi\}_t }\\
 &=
-{j \over m} \ip{\Psi}{ K^{\text{x}} \Psi }.
\end{split}
\]
\end{widetext}
We now use the formula
\begin{equation}
\label{eq mom mass vel}
\text{\it momentum} = \text{\it mass} \times \text{\it velocity}.
\end{equation}
Note that momentum and velocity are signed quantities, so that as long as mass is positive (\ref{eq mom mass vel}) stipulates that momentum and velocity have the same sign, i.e., direction.  We conclude that the hermitian operator
\[
P \eqdef - j K^\text{x}
\]
measures momentum.  

Consistency would require that we obtain the energy operator from the skew-hermitian generator of time-translation in the same way, i.e.,
\[
E = -j K^\text{t}.
\]
From (\ref{eq gen time-tran}) we obtain
\[
E = {1 \over 2m} K^\text{x}{}^2 = - {1 \over 2m} P^2,
\]
so $\text{\it energy} = - \text{\it momentum}^2 / 2m$.

Of course, this is not the usual formula for kinetic energy, but as we have discussed above, it is an acceptable formula as long as we adopt the corresponding convention for all physical systems.  It should be noted that inserting a minus sign in (\ref{eq schro}) changes nothing, as it is equivalent to replacing $j$ by the other imaginary unit.

The reason this discrepancy does not arise in conventional treatments is that time-translation by the amount $\tau$ is ordinarily defined in terms of the \emph{evolution} operator $E_\tau$, which acts on the instantaneous (Schr\"odinger) state according to the rule:  $E_\tau \psi$ is the state $\psi$ after it has been allowed to evolve for a time $\tau$, i.e.,
\[
E_\tau \Psi_t = \Psi_{t + \tau}.
\]
If we define $U'_\tau$ as the operator on global states given by
\[
\{ U'_\tau \Psi\}_t = E_\tau \Psi_t = \Psi_{t + \tau},
\]
then $U'_\tau = {U^{\text t}_\tau}^{-1}$, where $U^{\text t}$ is the time-translation operator defined above according to the same general rule as $U^{\text x}$, and
\begin{statement}
\label{stmnt time symm}
for any state $\Psi$ of a system $\system S$ and $\Theta$ of an observer $\system O$, $U_\tau \Psi$ appears to $\Theta$ as $\Psi$ appears to $U^\text{\normalfont t}_\tau \Theta$,
\end{statement}
just the opposite of Rule~\ref{stmnt standard symm}.

As noted above, in classical lagrangian mechanics, if we parameterize the history of a system by an arbitrary parameter $\tau$ and make time a function of $\tau$, the canonical momentum for the time coordinate is $-1$ times the usual energy.  It really is just historical accident that we assign the sign we do to energy.

\section{Time-reversal}
\label{sec time rev}

This section is devoted to justifying the assertion made in section~\ref{sec math struc} that time-reversal must ordinarily be conjugate-unitary.  A demonstration of this fact provides an illuminating exercise in the management of projective representations and conjugate-unitary operations, as well as some assurance that such operations are an essential feature of quantum mechanics, not a mathematical oddity.  Suppose therefore that a system $\system S$ enjoys time-reversal symmetry.  Let $\tau \mapsto A_\tau$ be the time-translation group of $\system S$, and let $\bs T$ be time-reversal.  Choose some fixed projective representation $\iota : \bs S \to \proj V$ of $\system S$.  Let $U_\cdot$ be a unitary representation of $A_\cdot$, and let $T$ be an $\eta$-unitary representation of $\bs T$ (via $\iota$).  Note that $\eta$ is uniquely determined by $\bs T$, i.e., by the physical operation of time-reversal, independently of the choice of $\iota$ (and $\vecsp V$, and $\mathbb C$).  Since $A_{-\tau} = \bs T \circ A_\tau \circ \bs T^{-1}$, $U_{-\tau} = \alpha_\tau T \circ U_\tau \circ T^{-1}$, where $\alpha_\tau$ is a phase factor.  For $\tau, \tau' \in \mathbb R$,
\[
\begin{split}
U_{-(\tau + \tau')}
 &=
U_{-\tau} U_{-\tau'}\\
 &=
\alpha_\tau T \circ U_\tau \circ T^{-1} \circ \alpha_{\tau'} T \circ U_{\tau'} \circ  T^{-1}\\
 &=
\alpha_\tau T \circ U_\tau \circ \eta(\alpha_{\tau'})  T^{-1} \circ T \circ U_{\tau'} \circ  T^{-1}\\
 &=
\alpha_\tau \eta\big(\eta(\alpha_{\tau'})\big)  T \circ U_\tau \circ U_{\tau'} \circ  T^{-1}\\
 &=
\alpha_\tau \alpha_{\tau'} T \circ U_{\tau + \tau'} \circ  T^{-1},
\end{split}
\]
so $\alpha_{\tau + \tau'} = \alpha_\tau \alpha_{\tau'}$, and with the usual assumptions of smoothness, for some $a \in \mathbb R$, $\alpha_\tau = \exp(i \tau a)$.  ($i$ is a fixed imaginary unit.  In this discussion issues related to the choice of imaginary unit and conjugacy class of statevector representation are not germane, and we revert to the customary name for the distinguished imaginary unit.)  Let $H$ be such that $U_\tau = \exp( i \tau H )$.  Then
\[
e^{-i \tau H} = e^{i \tau a} T e^{i \tau H} T^{-1}.
\]
Differentiating and evaluating at $\tau = 0$ we have
\[
-iH = ia\id + T i H T^{-1}
 = ia\id + \eta(i) T H T^{-1},
\]
so
\begin{equation}
\label{eq THT}
H = i\eta(i)  T H T^{-1} - a\id.
\end{equation}
(Note that $i \eta(i)$ is 1 or $-1$ according as $\eta$ is the conjugation map or the identity map on $\mathbb C$; in particular, it is real.)  

As discussed above, either $H$ or $-H$ measures the energy; it doesn't matter which.  Let $v \in V$ with $\|v\| = 1$ represent a physical state $\bs \sigma$.  The expectation of $H$ for $\bs \sigma$ is
\begin{equation}
\label{eq vHv}
\begin{split}
\ip{v}{Hv}
 &=
\ip{v}{i\eta(i) T H T^{-1} v - av}\\
 &=
i \eta(i) \eta \big( \ip{T^{-1} v}{H T^{-1}v} \big) - a\\
 &=
i \eta(i) \ip{T^{-1} v}{H T^{-1} v} - a,
\end{split}
\end{equation}
since $i \eta(i)$ is real, $T^{-1}$ is $\eta$-unitary, and $H$ is hermitian, so $\ip{T^{-1} v}{H T^{-1} v}$ is real.

$T^{-1} v$ represents $\bs T \bs \sigma$, and we see that the expectation of $H$ for $\bs T \bs \sigma$ is therefore $i \eta(i)$ times that for $\bs \sigma$, minus $a$.

If $\eta$ is the identity automorphism of $\mathbb C$ then $i \eta(i) = i^2 = -1$, so in this case time-reversal changes the sign of energy increments, which is impossible for any system whose energy spectrum is unbounded in one direction and not the other, which is essentially all real physical systems.

If $\eta$ is complex conjugation then $i \eta(i) = 1$, so from (\ref{eq THT}) we obtain
\[
T H T^{-1}
 =
H + a\id,
\]
and
\[
T^2 H T^{-2}
 =
H + 2 a\id.
\]
But $T^2$ is a multiple of the identity, and $T^{-2}$ is the reciprocal multiple of the identity, so $T^2 H T^{-2} = H$, and $a = 0$.  By virtue of (\ref{eq vHv}), the expectation of energy is therefore unchanged by time-reversal.

Note that we have not only shown that, for a system with time-reversal symmetry, time-reversal is represented by a conjugate-unitary operator, but also that the energy of a state is the same as that of the time-reversed state.

\section{Representations of compound systems}
\label{sec repr comp}

We have shown that the choice of conjugacy class of representation of physical systems is linked to the sign of certain physical quantities, including energy.  We have also shown that the choice of conjugacy class in a given case is just as arbitrary as---is indeed essentially equivalent to---the choice of imaginary unit in a complex field.  And we have shown that the standard choice of energy operator is the opposite of what it reasonably should be.  On the other hand, we have discussed the disastrous consequences of changing the conjugacy class of representation (and therefore the sign of energy) of any given system independently of another system with which it can interact.  The fact that ``all hell would break loose'' if we did so cannot be the only bar to this.  Moreover, \emph{ad hoc} rules for the choice of representation based on the spectra of generators of continuous symmetries, or on conservation principles derived from such symmetries, are useless for systems that do not enjoy the symmetries in question.  Is it possible that in such cases the conjugacy classes of representations of diverse systems are unlinked?  It hardly seems likely, and in this section we show how and why the conjugacy class of representation of any system is tied to that of any other system with which it may be meaningfully composed.

As discussed above, given a physical system $\system S = (\bs S, \mathbb R, \simil{\cdot}{\cdot} )$ and a SCIP space $\vecsp V = (V, \mathbb C, \mathbb R, \ip{\cdot}{\cdot} )$ such that $\system S$ is isomorphic to the projective space $\proj{\vecsp V}$, the sign of energy and all other physical symmetry generators is determined by the choice of conjugacy class of the isomorphism $\iota : \system S \to \proj{\vecsp V}$ and the choice of imaginary unit in $\mathbb C$.  Without loss of generality, therefore, we may replace the complex field $\mathbb C = (C, R, <, +, \cdot)$ by a complex field wdiu $(\mathbb C, j) = (C, R, <, +, \cdot, j)$, where $j$ is either imaginary unit for $\mathbb C$.  Complex fields wdiu are like real fields in that there is a unique isomorphism relating any two of them, so without loss of generality we may regard all complex fields wdiu, wherever they occur, as identical.  We will use `$i$' to denote the distinguished imaginary unit in this (these) field(s).  All the ambiguity now resides in the conjugacy class of the isomorphism used to obtain the projective representation of any given physical system.

Suppose $\system A$ and $\system B$ are quantum systems.  To avoid irrelevant complications, we suppose $\system A$ and $\system B$ to be distinguishable in the following sense.  Given particular states $\boldsymbol \alpha$ of $\system A$ and
$\boldsymbol \beta$ of $\system B$, a unique state $\boldsymbol \alpha\boldsymbol \beta$ of the compound system $\system E = \system A \system B$ is
determined.  Moreover, if either $\boldsymbol \alpha \ne
\boldsymbol \alpha'$ or $\boldsymbol \beta \ne \boldsymbol \beta'$
then $\boldsymbol \alpha\boldsymbol \beta \ne \boldsymbol
\alpha'\boldsymbol \beta'$.  In other words, there is a well
defined map,
\[
\boldsymbol \alpha, \boldsymbol \beta \mapsto
 \boldsymbol \alpha\boldsymbol \beta,
\]
from $\system A \times \system B$ to $\system E$, and this map is
one-to-one.

Recall that the similarity of physical states $\boldsymbol \sigma$ and $\boldsymbol \sigma'$ is the probability that the answer to the question `is the state $\boldsymbol \sigma'$?' is `yes' when the state is $\boldsymbol \sigma$.  It is a basic principle that questions of this form for $\system A$ and for $\system B$ are commuting quantum observables.  In general, any observation that asks only about $\system A$ commutes with any
observation that asks only about $\system B$.

It follows that if $\boldsymbol \varepsilon_1 = \boldsymbol \alpha_1\boldsymbol
\beta_1$ and $\boldsymbol \varepsilon_2 = \boldsymbol \alpha_2\boldsymbol
\beta_2$ are two elements of $\system E$ that happen to be pure
product states,
\begin{equation}
\label{ch}
\simil{\boldsymbol \varepsilon_1}{\boldsymbol \varepsilon_2}
 = \simil{\boldsymbol \alpha_1}{\boldsymbol \alpha_2}
    \simil{\boldsymbol \beta_1}{\boldsymbol \beta_2}.
\end{equation}

As before, we let $\vecsp A$, $\vecsp B$, and $\vecsp E$ be SCIP spaces, and let $\iota^A$, $\iota^B$, and $\iota^E$ be isomorphisms of $\system A$, $\system B$, and $\system E$ with their respective projective spaces.  As discussed in Section~\ref{sec math struc}, the respective real fields of these structures are related by unique isomorphisms, so we may suppose them all to be the same field without loss of generality.  Indeed, we have implicitly supposed this to be the case, in order that (\ref{ch}) be meaningful.  We refer to this common real field as $\mathbb R$.  As discussed just above, we may suppose the respective complex fields to be with distinguished imaginary unit, so that they too are related by unique isomorphisms, and we may also suppose these all to be the same field, which we refer to as $\mathbb C$.  Note that if we suppose at the outset that the vector spaces involved all incorporate the same complex field, there is no need to suppose that it has a distinguished imaginary unit.

The map $\boldsymbol \alpha, \bs \beta \mapsto \bs\alpha \bs\beta$ corresponds to a map $\boldsymbol T : \proj{\vecsp A} \times \proj{\vecsp B} \to \proj{\vecsp E}$ satisfying:
\begin{equation}
\label{eq prod proj}
\simil{\boldsymbol T(\boldsymbol \alpha_1,\boldsymbol \beta_1)}
   {\boldsymbol T(\boldsymbol \alpha_2,\boldsymbol \beta_2)}
 = \simil{\boldsymbol \alpha_1}{\boldsymbol \alpha_2}
   \simil{\boldsymbol \beta_1}{\boldsymbol \beta_2}.
\end{equation}
This basic identity, generalized to an arbitrary finite number of constituent systems, is the starting point for the following fundamental theorem.
\begin{theorem}
\label{ycH}
Suppose $\vecsp A_1$, $\vecsp A_2$, \dots, $\vecsp A_N$, and $\vecsp E$ are SCIP spaces of dimension at least 2, and suppose $\boldsymbol T : \proj{\vecsp A}_1
\times \dotsm \times \proj{\vecsp A}_N \to \proj{\vecsp E}$ satisfies
\begin{multline}
\label{eq prod proj n}
\simil{\boldsymbol T(\boldsymbol \alpha_1,\dots,\boldsymbol \alpha_N)}
   {\boldsymbol T(\boldsymbol \alpha'_1,\dots,\boldsymbol \alpha'_N)}\\
 = \simil{\boldsymbol \alpha_1}{\boldsymbol \alpha'_1}
 \dotsm \simil{\boldsymbol \alpha_N}{\boldsymbol \alpha'_N}.
\end{multline}
Then there is a map $T:\vecsp A_1 \times \dotsm \times \vecsp A_N \to \vecsp E$ that represents $\boldsymbol T$ and has the following two
properties.  First, $T$ is $\star$-linear in each of its arguments.  That is, there is an $N$-sequence $\boldsymbol \eta = \langle \eta_1, \dots, \eta_N \rangle$ of automorphisms of $\mathbb C$ (each individually either the identity or the conjugation map), such that $T$ is $\eta_n$-linear in its $n$th argument for $n = 1, \dots, N$.  Second, $T$ satisfies the $\boldsymbol \eta$-unitarity condition:
\begin{multline}
\label{yCH}
\ip{T(\alpha_1, \dots, \alpha_N)}{T(\alpha'_1, \dots, \alpha'_N)}\\
 = \eta_1(\ip{\alpha_1}{\alpha'_1})
 \dotsm \eta_N(\ip{\alpha_N}{\alpha'_N}).
\end{multline}
$\boldsymbol\eta$ is uniquely determined, and $T$ is determined up to a phase factor.
\end{theorem}
Note that the supposition of single complex field is required in order that (\ref{yCH}) make sense.

\begin{proof}The proof is by induction on $N$.  For $N = 1$, the theorem is just Wigner's theorem.  We now suppose the theorem to be true for $N = M$ and will prove it for $N = M+1$.  We therefore suppose that $\vecsp A_1$, $\vecsp A_2$, \dots, $\vecsp A_{N}$, and $\vecsp E$ are SCIP spaces, and that $\boldsymbol T : \proj{\vecsp A}_1 \times \dotsm \times \proj{\vecsp A}_{N} \to \proj{\vecsp E}$ satisfies (\ref{eq prod proj n}) with $N = M+1$.

Condition~\ref{yCH} may be replaced by the two conditions:
\begin{statement}
\label{Ch1}
If $\alpha_n \perp \alpha'_n$ for some $n \in \{ 1, \dots, N \}$ then $T(\alpha_1, \dots, \alpha_{N})\perp T(\alpha'_1, \dots, \alpha'_{N})$.
\end{statement}
\begin{statement}
\label{Ch2}
$\|T(\alpha_1, \dots, \alpha_{N})\| = \|\alpha_1\| \dotsm \|\alpha_{N}\|$.
\end{statement}
To show that these conditions suffice, suppose that $T$ is $\eta_n$-linear in its $n$th argument for each $n \in \{1, \dots, N\}$, where each $\eta_n$ is an automorphism of $\mathbb C$ (either the identity or the conjugation map)---and let $\alpha_n, \alpha'_n \in \vecsp A_n$ be given for all $n \in \{1, \dots, N\}$.  We want to evaluate
$\ip{T(\alpha_1, \dots, \alpha_{N})}{T(\alpha'_1, \dots, \alpha'_{N})}$, assuming $T$ satisfies Conditions~\ref{Ch1} and \ref{Ch2}.

For each $n \in \{1, \dots, N\}$, let $\theta_{n1}$ and $\theta_{n2}$ be an orthogonal pair of normalized vectors in $\vecsp A$ such that $\alpha_n = a_{n1}\theta_{n1} +
a_{n2}\theta_{n2}$, for some $a_{n1}, a_{n2}$, and $\alpha'_n = a'_{n1}\theta_{n1} +
a'_{n2}\theta_{n2}$, for some $a'_{n1}, a'_{n2}$.  (To avoid trivial cases, we have assumed that $\dim\vecsp A_n \ge 2$ for all $n$.)  Using the $\boldsymbol \eta$-linearity of $T$ we may write:
\begin{widetext}
\begin{align*}
T(\alpha_1, \dots, \alpha_N)
 &=
 \sum_{i_1 = 1}^2 \dotsm \sum_{i_N = 1}^2 \eta_1(a_{1i_1})
 \dotsm \eta_N(a_{Ni_N})T(\theta_{1i_1}, \dots, \theta_{Ni_N}),\\
T(\alpha'_1, \dots, \alpha'_N)
 &=\sum_{i_1 = 1}^2 \dotsm \sum_{i_N = 1}^2 \eta_1(a'_{1i_1})
 \dotsm \eta_N(a'_{Ni_N})T(\theta_{1i_1}, \dots, \theta_{Ni_N}).
\end{align*}

We therefore have:
\begin{align*}
\ip{T(\alpha_1, \dots, \alpha_N)}{T(\alpha'_1, \dots, \alpha'_N)}
 &=
\sum_{i_1 = 1}^2 \dotsm \sum_{i_N = 1}^2
  \overline{\eta_1(a_{1i_1}) \dotsm \eta_N(a_{Ni_N})}
  \eta_1(a'_{1i_1}) \dotsm \eta_N(a'_{Ni_N})\\
 &=
\Big(\sum_{i=1}^2 \overline{\eta_1(a_{1i})}\eta_1(a'_{1i})\Big)
 \dotsm
 \Big(\sum_{i=1}^2 \overline{\eta_N(a_{Ni})}\eta_N(a'_{Ni})\Big)\\
 &=
\eta_1(\ip{\alpha_1}{\alpha'_1})
 \dotsm
 \eta_N(\ip{\alpha_N}{\alpha'_N}),
\end{align*}
\end{widetext}
as claimed.

We now construct a map $T$ as desired.  It will be convenient to refer to $\vecsp V_N$ as $\vecsp B$, and to indicate sequences of length $M$ by an underline, with various conventions, such as: `$\underline \alpha_0$' to mean `$\alpha_{01}, \dots, \alpha_{0M}$',  `$\underline \alpha \in \underline {\vecsp A}$' to mean `$\alpha_n \in \vecsp A_n$ for $n = 1, \dots, M$', and `$\underline \alpha, \beta$' to mean `$\alpha_1, \dots, \alpha_M, \beta$'.  (Recall that $N = M + 1$.)

Let $\underline\alpha_0 \in \underline {\vecsp A}$ and $\beta_0 \in \vecsp B$ be arbitrary unit vectors.  Let $\underline{\bs \alpha}_0 = \underline{\proj\alpha}_0$ and $\bs
\beta_0 = \proj\beta_0$.  $T(\underline\alpha_0,\beta_0)$ may be taken to
be any unit vector $\theta_0$ in the ray $\bs
T(\bs \alpha_0,\bs \beta_0)$.  $T$ is uniquely
determined by the choice of $\theta_0$, as the following
construction will show.  Since $\theta_0$ is determined up to a
phase factor, the same will be true of $T$.

Define $\bs V : \proj{\vecsp A}_1 \times \dotsm \times \proj{\vecsp A}_M \to \proj{\vecsp E}$ by:
\[
\bs V(\underline{\bs \alpha})
 = \bs T(\underline{\bs \alpha},\bs \beta_0).
\]
By the induction hypothesis there are automorphisms $\underline\eta = \eta_1, \dots, \eta_M$ of $\mathbb C$ and an $\underline\eta$-unitary map $V:\vecsp A_1 \times \dotsm \times \vecsp A_M \to \vecsp E$ that represents $\bs V$, which is determined up to a phase factor.  We fix the phase by requiring $V(\underline \alpha_0) = \theta_0$.

For any $\underline{\bs \alpha} \in \underline{\proj{\vecsp A}}$ define
$\bs U_{\underline{\bs \alpha}}$ by:
\[
\bs U_{\underline{\bs \alpha}}(\bs \beta)
 =\bs T(\underline{\bs \alpha},\bs \beta).
\]
Given a nonzero $\underline \alpha \in \underline{\vecsp A}$, we apply Wigner's Theorem
to $\bs U_{\underline{\proj\alpha}}$ and obtain a $\star$-unitary
map $U : \vecsp B \to \vecsp E$ that represents $\bs
U_{\underline{\proj\alpha}}$.  Define $S_{\underline\alpha}$ to be that scalar multiple
of $U$ such that $S_{\underline\alpha}(\beta_0) = V(\underline \alpha)$.  Let $S_0
= 0$.  One can show by a topological or by an algebraic
argument that $S_{\underline\alpha}$ is either linear for all $\underline\alpha$ or
conjugate-linear for all $\underline\alpha$.  We define $\eta_N$ to be the
automorphism of $\mathbb C$ that is common to all the
$S_{\underline\alpha}$'s.

Finally, we define
\m{ci}
\begin{equation}
\label{ci}
T(\underline\alpha,\beta) = S_{\underline\alpha}(\beta).
\end{equation}
Our construction has guaranteed that for all $\underline\alpha \in \underline{\vecsp
A}$ and $\beta \in \vecsp B$, $\big(T(\underline\alpha,\beta)\big)\spproj =
\bs T(\underline{\proj\alpha},\proj\beta)$, so $T$ represents
$\bs T$.  From this it follows that Condition~\ref{Ch1}
is satisfied.  Since $\|\beta_0\| = 1$ by choice, and $V(\cdot)$ is $\star$-unitary,
\[
\begin{split}
\|T(\underline\alpha,\beta)\|
 &=
\|S_{\underline\alpha}(\beta)\|
 =
\|S_{\underline\alpha}(\beta_0)\| \|\beta\|
 =
\|V(\underline\alpha)\| \|\beta\|\\
 &=
\|\alpha_1\| \dotsm \|\alpha_M\| \|\beta\|,
\end{split}
\]
so Condition~\ref{Ch2} is satisfied.

It remains to be shown that $T(\cdot)$ is \mbox{($\underline \eta, \eta_N$)}-linear.  It is straightforward to
show that:
\[
T(\underline a \underline \alpha, b\beta) = \eta_1(a_1) \dotsm \eta_M (a_M) \eta_N(b) T(\underline\alpha,\beta).
\]
It is equally simple to show that $T$ is additive in its last argument, as
\[
T(\underline\alpha,\beta+\beta')
 = S_{\underline\alpha}(\beta+\beta')
 = T(\underline\alpha,\beta) + T(\underline\alpha,\beta').
\]
Proving additivity in each of the first $M$ arguments is more involved.  For notational convenience, we will carry out the proof for the first argument, but it clearly generalizes to any of the first $M$ arguments.  Let $\alpha_m \in \vecsp V_m$, $m \in \{ 2, \dots, M \}$, and $\beta \in \vecsp B$ be fixed.  Let `$\hat \alpha$' stand for `$\alpha_2, \dots, \alpha_M$', with related conventions homologous to those involving `$\underline\alpha$'.  In light of the definition\erf{ci} of $T$, we must show that for any $\alpha,\alpha' \in \vecsp A_1$,
\begin{equation}
\label{CJ}
S_{(\alpha + \alpha'), \hat \alpha}(\beta)
 = S_{\alpha, \hat\alpha}(\beta) + S_{\alpha', \hat \alpha}(\beta).
\end{equation}
Suppose for the moment that we have this identity for the case $\alpha \perp \alpha'$.  We may then derive the general formula as follows.  Given arbitrary $\alpha, \alpha' \in \vecsp A_1$, let $\theta_1, \theta_2 \in \vecsp A_1$ and $a_1, a_2, a'_1, a'_2 \in \mathbb C$ be such that $\theta_1 \perp \theta_2$, $\alpha = a_1 \theta_1 + a_2 \theta_2$, and $\alpha' = a'_1 \theta_1 + a'_2 \theta_2$.  Using the multiplicativity and orthogonal additivity properties (for multiples of $\theta_1$ and $\theta_2$) we have
\begin{widetext}
\[
\begin{split}
S_{(\alpha + \alpha'), \hat \alpha}(\beta)
 &=
S_{((a_1 + a'_1) \theta_1 + (a_2 + a'_2) \theta_2), \hat \alpha}(\beta)
 =
\eta_1(a_1 + a'_1)S_{\theta_1, \hat \alpha}(\beta)
 + \eta_1(a_2 + a'_2)S_{\theta_2, \hat \alpha}(\beta)\\
 &=
\eta_1(a_1) S_{\theta_1, \hat \alpha}(\beta)
 + \eta_1(a_2) S_{\theta_2, \hat \alpha}(\beta)
 + \eta_1(a'_1) S_{\theta_1, \hat \alpha}(\beta)
 + \eta_1(a'_2) S_{\theta_2, \hat \alpha}(\beta)\\
 &=
S_{(a_1 \theta_1 + a_2 \theta_2), \hat \alpha}(\beta)
 + S_{(a'_1 \theta_1 + a'_2 \theta_2), \hat \alpha}(\beta),
\end{split}
\]
\end{widetext}
as claimed.

All that's left is to prove orthogonal additivity, so suppose $\alpha \perp \alpha'$ and let $\alpha'' = \alpha + \alpha'$.  If either $\alpha$ or $\alpha'$ is 0, the result is trivial, so we henceforth assume that neither vanishes.  By the pythagorean property of the inner product, $\simil{\alpha''}{\alpha} + \simil{\alpha''}{\alpha'} = 1$.  It follows from Condition~\ref{Ch2} that
\begin{equation}
\label{eq pythag}
\begin{split}
&\simil{T(\alpha'',\hat \alpha,\beta)} {T(\alpha,\hat \alpha,\beta)}
   + \simil{T(\alpha'',\hat \alpha,\beta)} {T(\alpha',\hat \alpha,\beta)}\\
&\qquad\qquad= \simil{\alpha''}{\alpha} \simil{\hat \alpha}{\hat \alpha} \simil\beta\beta
   + \simil{\alpha''}{\alpha'} \simil{\hat \alpha}{\hat \alpha} \simil\beta\beta\\
&\qquad\qquad= \simil{\alpha''}{\alpha} + \simil{\alpha''}{\alpha'}
 = 1.
\end{split}
\end{equation}
By Condition~\ref{Ch1}, $T(\alpha,\hat \alpha,\beta) \perp T(\alpha',\hat \alpha,\beta)$, so $T(\alpha'',\hat \alpha,\beta)$ is a linear
combination of $T(\alpha,\hat \alpha,\beta)$ and $T(\alpha',\hat \alpha,\beta)$.  (By the pythagorean property of the inner product,
\[
\begin{split}
1
 &=
\simil{T(\alpha'',\hat \alpha,\beta)} {T(\alpha,\hat \alpha,\beta)}
 + \simil{T(\alpha'',\hat \alpha,\beta)} {T(\alpha',\hat \alpha,\beta)}\\
 &\phantom=\quad
 + \simil{T(\alpha'',\hat \alpha,\beta)} {\gamma},
\end{split}
\]
where $\gamma \perp T(\alpha,\hat \alpha,\beta), T(\alpha',\hat \alpha,\beta)$ is such that $T(\alpha'',\hat \alpha,\beta)$ is a linear combination of $T(\alpha,\hat \alpha,\beta)$, $T(\alpha',\hat \alpha,\beta)$, and $\gamma$.  If $T(\alpha'',\hat \alpha,\beta)$ were not a linear
combination of $T(\alpha,\hat \alpha,\beta)$ and $T(\alpha',\hat \alpha,\beta)$, the third term would be nonzero, violating (\ref{eq pythag}).)

In terms of the $S$ maps, we therefore have
\begin{equation}
\label{cI}
S_{\alpha'',\hat \alpha}(\beta)
 = aS_{\alpha,\hat \alpha}(\beta) + a'S_{\alpha',\hat \alpha}(\beta),
\end{equation}
for some $a,a'$.  We wish to show that
\begin{equation}
\label{cII}
S_{\alpha'',\hat \alpha}(\beta)
 = S_{\alpha,\hat \alpha}(\beta) + S_{\alpha',\hat \alpha}(\beta),
\end{equation}
If $\beta$ is proportional to $\beta_0$, the additivity we are seeking to prove is just the additivity of $V(\cdot)$, so we now assume that $\beta$ is not proportional to $\beta_0$.  Write $\beta$ in the form
\[
\beta = b\beta_0 + \beta_1,
\]
where $\beta_0 \perp \beta_1$ and $\beta_1 \neq 0$.

By the $\eta_N$-linearity of the $S$ maps, we have
\m{Ci}
\begin{equation}
\label{Ci}
S_{\alpha'',\hat \alpha}(\beta)
 = \eta_N(b)S_{\alpha'',\hat \alpha}(\beta_0)
   + S_{\alpha'',\hat \alpha}(\beta_1).
\end{equation}

Similarly,
\m{CI}
\begin{align}
\label{CI}
S_{\alpha,\hat \alpha}(\beta)
 = \eta_N(b)S_{\alpha,\hat \alpha}(\beta_0)
   + S_{\alpha,\hat \alpha}(\beta_1),\\
\intertext{and}
S_{\alpha',\hat \alpha}(\beta)
 = \eta_N(b)S_{\alpha',\hat \alpha}(\beta_0)
   + S_{\alpha',\hat \alpha}(\beta_1).
\end{align}

Replacing `$\beta$' in (\ref{cI}) by `$\beta_1$', we have:
\m{cJ}
\begin{equation}
\label{cJ}
S_{\alpha'',\hat \alpha}(\beta_1)
 = a_1S_{\alpha,\hat \alpha}(\beta_1) + a_1'S_{\alpha',\hat \alpha}(\beta_1),
\end{equation}
for appropriate $a_1$ and $a_1'$.  By the definition of the $S$
maps
\m{cj}
\begin{equation}
\label{cj}
\begin{split}
S_{\alpha'',\hat \alpha}(\beta_0) = V(\alpha'',\hat \alpha)
 &= V(\alpha,\hat \alpha) + V(\alpha',\hat \alpha)\\
  &= S_{\alpha,\hat \alpha}(\beta_0) + S_{\alpha',\hat \alpha}(\beta_0).
\end{split}
\end{equation}
Substituting (\ref{cI}), (\ref{CI}), (\ref{cJ}), and (\ref{cj})
in (\ref{Ci}), we obtain:
\begin{multline*}
\eta_N(b)(S_{\alpha,\hat \alpha}(\beta_0) + S_{\alpha',\hat \alpha}(\beta_0))\\
  + a_1S_{\alpha,\hat \alpha}(\beta_1) + a_1'S_{\alpha',\hat \alpha}(\beta_1)\\
 = a(\eta_N(b)S_{\alpha,\hat \alpha}(\beta_0) + S_{\alpha,\hat \alpha}(\beta_1))\\
  + a'(\eta_N(b)S_{\alpha',\hat \alpha}(\beta_0) + S_{\alpha',\hat \alpha}(\beta_1)).
\end{multline*}
Since $S_{\alpha,\hat \alpha}(\beta_0)$, $S_{\alpha,\hat \alpha}(\beta_1)$,
$S_{\alpha',\hat \alpha}(\beta_0)$, and $S_{\alpha',\hat \alpha}(\beta_1)$ are mutually orthogonal and nonzero (by virtue of our assumptions that $\alpha$, $\alpha'$, $\alpha_2$, \dots, $\alpha_M$, $\beta_0$, and $\beta_1$ do not vanish), they are independent, so we can match the coefficients on the left and right sides above to obtain in particular $\eta_N(b) =
a\eta_N(b)$ and $\eta_N(b) = a' \eta_N(b)$.  If $\beta$ is not orthogonal to $\beta_0$ then $b \neq 0$, so $a = a' = 1$, and (\ref{cI}) is the desired identity.

To handle the case that $\beta$ is orthogonal to $\beta_0$ we simply write $\beta = \beta' + \beta''$ where $\beta'$ and $\beta''$ are not orthogonal to $\beta$.  Since the $S$ maps are linear, (\ref{cII}) for $\beta$ follows from (\ref{cII}) for $\beta'$ and $\beta''$.\qed
\end{proof}

\section{Compatibility of representations}
\label{sec comp repr}

It is important to remember that the analysis in the preceding section is predicated on the assumption that the complex fields incorporated in the spaces $\vecsp A_1$, \dots, $\vecsp A_N$, and $\vecsp E$ are either all the same field, or are fields with distinguished imaginary unit.  In the latter case, the existence of unique isomorphisms relating the fields renders it equivalent to the former case, with the added feature of a distinguished imaginary unit.  We will henceforth suppose that all vector spaces incorporate the same complex field, with or without a distinguished imaginary unit.  As we saw in Section~\ref{sec con sym}, a choice of imaginary unit is required to determine the sign of generators of continuous symmetry groups, such as the energy operator.

We have noted above that the choices of conjugacy classes of representations of several systems $\system A_1, \dots, \system A_N$ may be expected to be dependent if the systems ``may be meaningfully combined''.  It is now clear that this should mean that the composite system $\system E = \system A_1 \dotsm \system A_N$ satisfies the superposition principle, so that $\system E$ is isomorphic to the full projective space of a SCIP space, not just to the subset of rays corresponding to pure-product states.  Theorem~\ref{ycH} depends on this.  We say that the systems are \emph{composable} in this case.

Suppose now that $\system A_1, \dots, \system A_N$ are composable, and let $\system E = \system A_1 \dotsm \system A_N$, with $\system E$ isomorphic to $\proj{\vecsp V}$ and $\system A_n$ isomorphic to $\proj{\vecsp V}_n$ for each $n$.  Given states $\bs \alpha_n$ and $\bs \alpha'_n$ of $\system A_n$ for each $n$, if we let $\bs \epsilon = \bs \alpha_1 \dotsm \bs \alpha_N$ and $\bs \epsilon' = \bs \alpha'_1 \dotsm \bs \alpha'_N$,
\[
\simil{\bs \epsilon}{\bs \epsilon'}
 =
\simil{\bs \alpha_1}{\bs \alpha'_1} \dotsm \simil{\bs \alpha_N}{\bs \alpha'_N}.
\]
Theorem~\ref{ycH} tells us that there is a map $T : \vecsp A_1 \times \dotsm \times \vecsp A_N \to \vecsp E$ that is $\star$-unitary in each of its arguments, such that if $\alpha_n$ represents $\bs \alpha_n$ for each $n$, $T(\alpha_1, \dots, \alpha_N)$ represents $\bs \alpha_1 \dotsm \bs \alpha_N$.  Of course, $T$ depends on the particular statevector representations we have chosen for $\system A_1$, \dots, $\system A_N$, and $\system E$, but by changing, if necessary, the conjugacy class of one or more of these representations, we can arrange that $T$ be unitary in all its
arguments.  There are exactly two ways of doing this, each being obtained from the other by reversing all the conjugacies at once.

Recall that the tensor product of vector spaces $\vecsp A_1, \dots, \vecsp A_N$ is a vector space $\vecsp E$ and a non-degenerate
multilinear map $T:\vecsp A_1 \times \dotsm \times \vecsp A_N \to \vecsp E$, such that the image of $T$ spans $\vecsp E$.  We write `$\vecsp A \otimes \dotsm \otimes \vecsp A_N$' for `$\vecsp E$', and `$\alpha_1 \otimes \dotsm \otimes \alpha_N$' for `$T(\alpha_1, \dots, \alpha_N)$'.  If the $\vecsp A_n$s are inner product spaces, we further require that $T$ be unitary in each of its arguments.

Theorem~\ref{ycH} therefore says that if $\system A_1$, \dots, $\system A_N$ are composable distinguishable quantum systems, representable in the SCIP spaces $\vecsp A_1$, \dots, $\vecsp A_N$, there are exactly two choices of conjugacy classes for the representations $\iota_n : \system A_n \to \vecsp A_n$ such that there exists a representation $\iota : \system A_1 \dotsm \system A_N \to (\vecsp A_1 \otimes \dotsm \otimes \vecsp A_N)\spproj$ of the compound system in the tensor product such that $\bs \alpha_1 \dotsm \bs \alpha_N$ is represented by $\alpha_1 \otimes \dotsm \otimes \alpha_N$.

We will say that representations $\iota_A$ and $\iota_B$ of composable systems $\system A$ and $\system B$ are \emph{compatible} iff they are consistent with a tensor product representation in this sense.  In this way a choice of representation for a given system $\system A$ determines a choice of conjugacy class of representations for all systems $\system B$ with which it is composable.  The relationship of composability of physical systems in quantum mechanics is surely quite general, perhaps universal.  Interactions of quantum systems typically take pure-product states to mixed states, so if $\system A$ and $\system B$ interact---or if there is any possibility of interaction---they are composable.  As discussed in the introduction, if there is no such possibility there is no reason a choice of representation of $\system A$ should constrain the choice of representation of $\system B$.

It is important that the notion of compatibility we have defined be an equivalence relation; otherwise it might impose inconsistent constraints on representations.  We obtain the reflexivity property by stipulating that any representation is compatible with itself.  (This is just a definition for the sake of completeness.  It has nothing to do with composing a system with itself, which is not meaningful in the context of this discussion.)  The symmetry condition is trivially satisfied.  It remains only to show the transitivity property, i.e., if $\iota_A$ is compatible with $\iota_B$, and $\iota_B$ is compatible with $\iota_C$, then $\iota_A$ is compatible with $\iota_C$.  An examination of the proof of Theorem~\ref{ycH} for the case $N=3$ shows that this is true.

\section{Conclusion}
\label{sec concl}

Suppose we stumble across a quantum system $\system A$ lying in the road, we pick it up and examine it, and we determine that as a similarity space $\system A = (\boldsymbol A, \mathbb R, \ip{\cdot}{\cdot})$ it is isomorphic to a SCIP space $\vecsp A = (A, \mathbb C, +_v, \cdot_s, \ip{\cdot}{\cdot})$ via some isomorphism $\iota^A$; if we have not already chosen an imaginary unit, any isomorphism---i.e., representation---is as good as any other.  To work this system into the rest of physics as we know it, however, we must choose an imaginary unit.  Which one to choose?  Our main result is that there is always exactly one choice that renders $\iota^A$ compatible with all the representations we are already using for other systems with which $\system A$ is composable, in the sense that for any such representation $\iota^B : \system B \to \vecsp B$, there is a representation $\iota : \system A \system B \to \vecsp A \otimes \vecsp B$ such that for all $\bs \alpha$ and $\bs \beta$, $\iota(\bs \alpha \bs \beta) = \iota^A(\bs \alpha) \otimes \iota^B(\bs \beta)$.  Alternatively, if our complex field has an imaginary unit at the outset, then the conjugacy class of $\iota^A$ is constrained by the compatibility requirement.  The constraint is imposed solely by the similarity structure of $\system A \system B$ \visavis those of $\system A$ and $\system B$.

Only after we have employed the compatibility requirement to fix the conjugacy class of the representation of $\system A$ may we proceed to use the behavior of $\system A$ under various continuous transformation groups such as time- and space-translation to define the corresponding observables---energy, momentum, etc.---from the skew-hermitian generators of the action of these groups on $\vecsp A$.  The signs of these observables are thereby fixed for $\system A$ relative to any of the various systems composable with it.  We may change all these signs at once, but we may not change any of them individually.

\section{Appendix}

Recall that a 1-dimensional physical symmetry group corresponds to a map $\mathbb R \ni \tau \mapsto A_\tau$ that is a homomorphism of the additive group of the reals into the group of automorphisms of the projective space $\proj V$ of a SCIP space $\vecsp V$.  As above, we make the physically reasonable assumption that $A_\cdot$ is sufficiently smooth that the conjugacy class of $A_\tau$ cannot change (from unitary to conjugate-unitary) discontinuously and therefore cannot change at all, so for every $\tau$, the operator representations of $A_\tau$ are unitary.  To obtain the usual mathematical setting of quantum mechanics, we must show that there exists a sufficiently well behaved group $\tau \mapsto U_\tau$ of unitary operators on $\vecsp V$, such that for each $\tau \in \mathbb R$, $\proj U_\tau = A_\tau$.

The following two theorems provide a definition of ``sufficiently good behavior''.  The first is due to Stone\cite{Stone:1932}. 
\begin{theorem}
Suppose $\tau \mapsto U_\tau$ is a 1-parameter group of operators on a Hilbert space $\vecsp H$, and suppose $U_\cdot$ is \emph{strongly continuous}, i.e., for any $u \in |\vecsp H|$ and $\tau_0 \in \mathbb R$, $\lim_{\tau \to \tau_0} U_\tau u = U_{\tau_0} u$.  Then there is a selfadjoint operator $K$ on $\vecsp H$, such that for all $\tau \in \mathbb R$, $U_\tau = \exp(\tau K)$.  For $u \in \dom L$,
\[
Ku = \left. {d U_\tau u \over d\tau} \right|_{\tau = 0},
\]
and the derivative exists iff $u \in \dom K$.
\end{theorem}
The second is due to von Neumann\cite{vonNeumann:1932}.
\begin{theorem}
Suppose $\tau \mapsto U_\tau$ is a 1-parameter group of operators on a Hilbert space $\vecsp H$, and suppose that for all $u, v \in |\vecsp H|$, the function $\tau \mapsto \ip{U_\tau u}{v}$ is measurable.  Then $U_\cdot$ is strongly continuous.
\end{theorem}
If we are not concerned with ``good behavior'' we can show the existence of such a group quite easily (assuming the axiom of choice).  By Zorn's lemma (the version of the axiom of choice most immediately applicable here) there exists a maximal set $T$ of real numbers such that the equation $\sum_{n=1}^N a_n \tau_n = 0$ cannot be satisfied with $N$ finite, $\tau_1, \dots, \tau_N$ distinct members of $T$, and $a_1, \dots, a_n$ integers not all 0.  Any real number then has a representation $\sum_{n=1}^N a_n \tau_n$, with $\tau_1$, \dots, $\tau_N$ distinct elements of $T$, which is unique up to permutation.  For  each $\tau \in T$ we let $U_\tau$ be such that $\proj U_\tau = A_\tau$ (using the axiom of choice again).  For $\tau = \sum_{n=1}^N a_n \tau_n$, we let $U_\tau = \prod_{n=1}^N U_{\tau_n}^{a_n}$.

This construction is essentially useless because it does not provide any mechanism to insure that the resulting representation is measurable in the sense of von Neumann's theorem.  Indeed, we may deliberately choose the operators $U_\tau$, $\tau \in T$ so that the group $\tau \mapsto U_\tau$ is quite pathological.  The same sort of construction, by the way, can be used to obtain groups $\tau \mapsto A_\tau$ of projective automorphisms that are equally pathological, which, presumably, actual physical symmetries are not.  The following heuristic argument suggests that a sufficiently smoothprojective group has a sufficiently well behaved linear representation.  In practice the existence of smooth linear representations is assumed at the outset, as this appears to model physical reality.  We use the term `smooth' informally; it generally corresponds to `differentiable' in a suitable sense.

To infer the existence of a well behaved linear representation $\tau \mapsto U_\tau$ we must assume sufficient smoothness of the projective group $\tau \mapsto A_\tau$.  Specifically, we assume that for any $r \in \proj V$, the function $\tau \mapsto A_\tau (r)$ is smooth at $\tau = 0$ in the sense that there exist a nonzero $u \in r$ and a $v \in \vecsp V$ (we will casually use `$\vecsp V$' for $|\vecsp V|$) such that
\begin{equation}
\label{eq smooth}
{d^2 \over d\tau^2} \simil{A_\tau (r)}{u + \tau v} = 0.
\end{equation}
The second derivative occurs here because $\simil r{r'}$ is in effect 1 minus the square of the ``distance'' between the rays $r$ and $r'$.  If we used the first derivative in (\ref{eq smooth}), the fact of its vanishing would not have much import as $\simil r{r'}$ always attains its maximum value (which is 1) when $r = r'$, so its derivative, if it exists, vanishes.

It is convenient to require that $v$ be orthogonal to $u$ and to specify that $\|u\| = 1$.  A simple computation shows that for any $v_0, v_1$ orthogonal to such a normalized $u$,
\[
{d^2 \over d\tau^2} \simil{u + \tau v_0}{u + \tau v_1}
 =
-2\|v_0 - v_1\|^2,
\]
from which it follows that (\ref{eq smooth}) can hold for at most one $v$ orthogonal to $u$.

For each $\tau \in \mathbb R$, a unitary representative $U_\tau$ of $A_\tau$ is specified by giving the value of $U_\tau u$.  Since $\| u + \tau v \|$ is constant to first order at $\tau = 0$, we may set $U_\tau u = u + \tau v$ for ``infinitesimal'' $\tau$, and extend this specification to all $\tau$ by the group property.  We define a partial linear operator $K$ by
\[
Kw = \left. {d U_\tau w \over d\tau} \right|_{\tau = 0},
\]
if this derivative exists; otherwise $w \notin \dom K$.  Note that $u \in \dom K$ and $Ku = v$.  Using the group property again we find that for any $\tau \in \mathbb R$, $U_\tau u \in \dom K$ and $K U_\tau u = U_\tau v$.  For $w \in \dom K$, let
\[
U'_\tau w = e^{\tau K}w.
\]
Then $U'_\tau$ and $U_\tau$ agree on $\dom U' = \dom K$.  If `God's in his Heaven - [and] All's right with the world'\cite{Browning:1841} (which is, after all, the essential heuristic assumption, refreshing in its na\"ivet\'e), $\dom K$ is dense in $\vecsp V$, and $U$ is the unique continuous extension of $U'$ to $\vecsp V$.

\end{document}